\def\lae{\mathrel{<\kern-1.0em\lower0.9ex\hbox{$\sim$}}}
\def\gae{\mathrel{>\kern-1.0em\lower0.9ex\hbox{$\sim$}}}

\newcommand{\be}{\begin{equation}}
\newcommand{\ee}{\end{equation}}

\documentclass[apj]{emulateapj}
\usepackage{rotating}
\usepackage{amsmath}
\shorttitle{Discerning the $\gamma$-ray emitting region} \shortauthors{Zheng et al.}

\begin{document}

\title{Discerning the $\gamma$-ray emitting region in the flat spectrum radio quasars}
\author{Y. G. Zheng\altaffilmark{1}; C. Y. Yang\altaffilmark{2,3}; L. Zhang\altaffilmark{4}; J.C. Wang\altaffilmark{2,3};}
\altaffiltext{1}{Department of Physics, Yunnan Normal University, Kunming, 650092, China (E-mail:ynzyg@ynu.edu.cn)}
\altaffiltext{2}{Yunnan Observatories, Chinese Academy of Sciences, Kunming 650011, China  (E-mail:chyy@ynao.ac.cn)}
\altaffiltext{3}{Key Laboratory for the Structure and Evolution of Celestial Objects, Chinese Academy of Sciences }
\altaffiltext{4}{Department of Astronomy, Yunnan University, Kunming, 650091, China }

\begin{abstract}
A model-dependent method is proposed to determine the location of the $\gamma$-ray emitting region for a given flat spectrum radio quasar (FSRQ). In the model, the extra-relativistic electrons are injected at the base of the jet and non-thermal photons are produced by both synchrotron radiation and inverse-Comtpon (IC) scattering in the energy dissipation region. The target photons dominating inverse-Comtpon scattering originate from both synchrotron photons and external ambient photon fields, and the energy density of external radiation field is a function of the distance between the position of dissipation region and a central super-massive black hole, and their spectra are seen in the comoving frame. Moreover, the energy dissipation region could be determined by the model parameter through reproducing the $\gamma$-ray spectra. Such a model is applied to reproduce the quasi-simultaneous multi-wavelength observed data for 36 FSRQs. In order to define the width of the broad-line region shell and dusty molecular torus shell, a simple numerical constraint is used to determine the outer boundary of the broad-line region and dusty molecular torus. Our results show that 1) the $\gamma$-ray emitting regions are located at the range from 0.1 pc to 10 pc; 2) the $\gamma$-ray emitting regions are located outside the broad-line regions and within the dusty molecular tori; and 3) the $\gamma$-ray emitting region are located closer to the dusty molecular torus ranges than the broad-line regions. Therefore, it may concluded that a direct evidence for the \emph{far site} scenario could be obtained on the basis of the model results.

\end{abstract}





\section{Introduction}

The powerful $\gamma$-ray emission, arising from the jet emission taking place in an radio-loud active galactic nuclei (AGN) whose jet axis is closely aligned with the observer's line of sight (Urry \& Padovani 1995), is a distinctive feature of a flat spectrum radio quasar (FSRQ). It is generally accepted that the $\gamma$-ray photons with energy above 100 MeV are probably attributed to the inverse-Compton scattering (ICS) off external ambient photon fields. There are two kinds of origins describing the target photons for ICS in the FSRQs, the ultraviolet photons from the gas in the broad-line region (BLR; e.g. Sikora et al. 1994; Fan et al. 2006) and/or the infrared photons from the dusty molecular torus (MT; Blazejowski et al. 2000; Arbeiter et al. 2002; Sokolov \& Marscher 2005), essentially resulting to both \emph{near site} and \emph{far site} of the $\gamma$-ray emission regions (e.g. Dotson et al. 2012). In the \emph{near site} scenario, the electron energy is dissipated inside the broad-line region (BLR) (e.g. Ghisellini \& Madau 1996; Georganopoulos et al. 2001) that locates at distances of $<$ 0.1-1 parsec (pc) from a central super-massive black hole (SMBH), while the \emph{far site} scenario argues that the electron energy is dissipated at several pc far from the SMBH (e.g. Lindfors et al. 2005; Sokolov \& Marscher 2005; Marscher et al. 2008), where the dominating population of the target photons will be from the MT, and the jet starts to be visible at the millimeter wavelength.

A large area telescope onboard the \emph{Fermi} (\emph{Fermi}-LAT) has shown the evidences for the high energy $\gamma$-ray emissions produced in the frame of \emph{near site} (Atwood et al. 2009). It has detected flares with variability timescale $\sim10^{4}$ s in some FSRQs (e.g. Abdo et al. 2009a; Abdo et al. 2010a; 2010b; Ackermann et al. 2010; Tavecchio et al. 2010; Foschini et al. 2011). Assuming that the entire cross section of the jet is emitting, above variability timescales would determine a dissipation region with a scale $\lesssim0.1$ pc. Alternatively, the \emph{near site} scenario is possible to explain the sharp breaks at GeV seen in the $\gamma$-ray spectra of some FSRQs by the opacity to pair production (Liu \& Bai 2006; Liu et al. 2008a; Bai et al. 2009; Poutanen \& Stern 2010; Stern\& Poutanen 2011). However, the sub-parsec scale energy dissipation is challenged by multi-wavelength simultaneous observations and polarimetry. In several cases, the optical polarimetry during an optical and $\gamma$-ray flare shows the polarization behavior similar to that observed in simultaneous very long baseline interferometry (VLBI) (Marscher et al. 2008; 2010; Abdo et al. 2010a; Jorstad et al. 2010; Agudo et al. 2011a). These facts indicate that both the bulk of $\gamma$-ray emissions are produced at further distances, even at distances of the order of 10-20 pc from the SMBH (Larionov et al. 2008; Sikora et al. 2008), and the $\gamma$-ray and VLBI jet emission sites seem to be co-spatial (e.g. Marscher et al. 2010; Jorstad et al. 2010; Agudo et al. 2011a; 2011b).

As an open issue, the origin of external ambient photon fields dominating the ICS in FSRQ jets should be traced. Since the target photons can be determined by the location of the $\gamma$-ray emitting region in the jet (e.g. Ghisellini \& Tavecchio 2009; Sikora et al. 2009; Ghisellini et al. 2010a; Agudo et al. 2011a), we argue that the scale of the energy dissipation region should be a clue on the ambient photon fields. There are two main diagnostics in the literature for the energy dissipation region: (i) the variability (e.g. Jorstad et al. 2010; Abdo et al. 2010c; Tavecchio et al. 2010; Liu et al. 2011a; 2011b; Agudo et al. 2012a; 2012b; Grandi et al. 2012; Brown 2013; Ramakrishnan et al. 2015), and (ii) the spectral energy distribution (SED) (e.g. Dermer et al. 2009; Ghisellini \& Tavecchio 2009; Zdziarski et al. 2012; Georganopoulos et al. 2012; Kang et al. 2015). Because the variability argument only implies a small dissipation region but not at any particular location (e.g. Giannios et al. 2009), the both the nature of target photons that depends on the accretion disc model (e.g. Shakura \& Sunyaev 1973; Narayan \& Yi 1994; 1995) and the electron energy dissipation location that is connected to the jet formation and collimation process (Vlahakis \& K$\rm \ddot{o}$nigl 2004; Marscher et al. 2008; Malmrose et al. 2011) should be take into account for the SED argument. In these scenarios, the issue of the $\gamma$-ray emitting region remains open.

In order to trace the origin of external ambient photon fields in the FSRQ's jets, in this paper, we propose a model-dependent method to determine the location of the $\gamma$-ray emitting region. In \S 2, we characterize the model; in \S 3, we give some numerical results, focusing on the location of the $\gamma$-ray emitting region;  in \S 4, we describe the sample; in \S 5, we model the distance between the dissipation region position and the SMBH on the {\bf basis} of reproducing the SEDs of a FSRQ sample; and in \S 6, we give the conclusion and discussion. Throughout the paper, we assume the Hubble constant $H_{0}=75$ km s$^{-1}$ Mpc$^{-1}$, the matter energy density $\Omega_{\rm M}=0.27$, the radiation energy density $\Omega_{\rm r}=0$, and the dimensionless cosmological constant $\Omega_{\Lambda}=0.73$.

\section{The Model}
The expected photon spectra in the context are produced by the model within the lepton model frame through both synchrotron radiation and ICS. In the model, we basically follow the approach of Potter \& Cotter (2012) to calculate the length evolution of the electron spectrum along with a stationary jet structure, and these electrons produce nonthermal photons through synchrotron radiation. We assume a length-dependent external ambient field which includes both BLR and MT for the ICS (Zheng \& Yang 2016). The model is characterized by following setups.

\subsection{Geometry of the jet structure}
In the model, the relativistic plasma propagates with an associated bulk Lorentz factor $\Gamma$ in a stationary funnel whose structure is constant with time in the lab frame, and the geometry is a truncated cone of the length $\Gamma L$ where the length of the jet in the lab frame is related to that in the fluid frame by a simple Lorentz contraction. We define the dynamic variable $x$ as the length along with the jet axis in the fluid frame, where $x=0$ is the base of the jet and $x=L$ is the end of the jet, so we can parameterize the geometry of the jet as follow:
\begin{equation}
R(x) = R_{0} + x \tan \theta_{\mathrm{open}} \;,
\label{Eq:1}
\end{equation}
where $R(x)$ is the radius of the jet at the length $x$, $R_{0}$ is radius at the base of the jet, and $\theta_{\mathrm{open}}$ is a half opening angle of the cone. We note that the jet opening angle $\theta^\prime_{\rm open}$ in the lab frame is related to the fluid frame opening angle $\theta_{\rm open}$ via $\Gamma \tan \theta^\prime_{\rm open}=\tan \theta_{\rm open}$. The stationary jet structure and magnetic energy conservation in each segment is determined by Eq. (\ref{Eq:1}) could have a relation:
\begin{equation}
B(x)=B_0\frac{R_0}{R(x)}\;,
\end{equation}
where $B(x)$ is the magnetic field of the jet at the length $x$, and $B_0$ is the magnetic field at the base of jet. The relation describes a pure toroidal field distribution in which the magnetic flux is parallel to the jet axis that increases with $x$.
\subsection{Energy equipartition}
It is well known that both the particle energy and field energy dominate the total energy of the jet plasma (e.g. Celotti \& Fabian 1993; Celotti et al. 2007; Finke et al. 2008). The energy equipartition could provide a minimum power solution on the emissions from the blazar jet. In the black hole-jet system, the relation between the particle energy and field energy depends on the uncertain jet formation, particle acceleration, and radiation mechanisms (e.g. Dermer et al. 2014). However, the baryon composition of the system is poorly known now (e.g. Reynolds et al. 1996; Kataoka et al. 2008; Ghisellini 2012; Kino et al. 2012). In our model, we assume that a condition of the equipartition holds between the magnetic field energy density $U_{B}$ and the non-thermal electron energy density $U_{e}$. In order to establish a parameter connection between the $U_{B}$ and $U_{e}$, we assume a equipartition fraction $A_{\rm equi}$,
\begin{equation}
A_{\rm equi}=\frac{U_B}{U_e}\;.
\end{equation}
From this equipartition fraction, a relation between $B_0$ and $R_0$ is found as (Potter \& Cotter 2012)
\begin{equation}
R_0=\sqrt{\frac{2E_jA_{\rm{equi}}\mu_0}{\Gamma^{2}(\pi B_0^2)(1+A_{\rm {equi}})}},
\end{equation}
where $E_j=P_{j}/c$ is the energy contained in a section of plasma with the width of 1 m and $P_{j}$ is the total jet power in the $x$-direction in the lab frame.

\subsection{Electron evolution}
The present model is to evolve the electron population dynamically along with the jet by taking into account energy losses from synchrotron emission and ICS. We consider the electron population in a slab with the width of 1 meter in the comoving frame and the jet structure is relative motion to the slab. Since a section of the jet with the width of $dx$ at any length $x$ containing $n$ meter slabs travels with the velocity of light $c$ towards the slabs, a slab takes $n/c$ seconds to cross the section in the fluid frame. It is noted that the electrons in a slab lose an amount of the energy which is equal to the total power $P_{tot}(x,dx,E_e)$ emitted by the section in the comoving frame divided by $c$ (Potter \& Cotter 2012). In this scenario, the evolution of the electron population along with the jet due to energy losses can be found by solving:
\begin{equation}
N_e(E_e,x+dx)=N_e(E_e,x)-\frac{P_{\rm tot}(x,dx,E_e)}{cE_e}\;,
\label{Eq:5}
\end{equation}
where $E_e$ is the electron energy in comoving frame. Assuming that electrons emit all of their energy at a critical frequency $\nu_{\rm cr}(x)=3eE_{e}^{2}B(x)/4\pi m_{e}^{3}c^{5}$, we can obtain
\begin{equation}
P_{\rm tot}(x,dx,E_e)=P_{\rm syn}(x,dx,E_e)+P_{\rm ic}(x,dx,E_e)\;,
\label{Eq:6}
\end{equation}
where $P_{\rm syn}(x,dx,E_e)$ is the synchrotron emission power and $P_{\rm ic}(x,dx,E_e)$ is the ICS power corresponding to a critical energy $E_{\rm cr}(x)=h\nu_{cr}(x)$ at the jet length $x$. In order to solve Eq.(\ref{Eq:5}), a initial injected electron distribution with the cut off energy $E_{\rm e,cut}$ has been assumed,
\begin{equation}
N_e(E_e,0)\simeq A_{0}E_e^\alpha \rm{e}^{-E_e/E_{\rm e, cut}}\;,
\end{equation}
where, $A_{0}\simeq(2-\alpha)E_{j}/[\Gamma^{2}(1+A_{\rm equi})(E_{\rm e,cut}^{2-\alpha}-E_{e,min}^{2-\alpha})]$ (Potter \& Cotter 2012) with the initial injected electron minimum energy $E_{e,min}$ , and $\alpha$ is electron spectral index.

\subsection{Synchrotron photon fields}
In order to produce the synchrotron photons through the jet fluid, we expect to know the opacity of the plasma. Assuming that the photons are observed at an angle smaller than the opening angle of the jet in the lab frame, Potter \& Cotter (2012) argued that the observer sees through to the layer within the opening angle where the total column optical depth is approximately 1 and beyond this there is an exponentially decreasing contribution from further layers. This suggests that the total absorption optical depth $\tau_{\rm tot}(E_{s},x)$ is the summation of the optical depths from each segment. In this scenario, we could obtain the $\tau_{\rm tot}(E_{s},x)$ in the comoving frame using the Lorentz transform,
\begin{equation}
\tau_{\rm tot}(E_{s},x)=\int^{L}_{x}k(E_{s},x)\Gamma^{2}(\frac{1}{\cos\theta_{\mathrm{observe}}}-\beta)dx\;,
\end{equation}
where $E_{s}$ is the energy of the synchrotron photon, $k(E_{s},x)$ is the synchrotron absorption coefficient at the jet length $x$, $\theta_{\mathrm{observe}}$ is the angle of observer's line of sight to the jet axis in the lab frame, and $\beta$ is the speed of jet material in the unit of $c$. Introducing a fully integration with the modified Bessel functions of order 5/3 to substitute for an approximate single energy emitting in Eq. (\ref{Eq:6}), we write the  synchrotron emission photon producing rate by an individual segment with the width $dx$ in the comoving frame as.
\begin{eqnarray}
w_{\rm syn}(x,dx,E_{s})&=&\frac{\sqrt{3}e^{3}B(x)}{h m_{e}c^{2}}\int_{E_{e,min}}^{E_{e,cut}}N_e(E_e,x)\nonumber\\&\times&F\left[\frac{E_{s}}{E_{\rm cr}(x)}\right]dE_e\;,
\label{Eq:9}
\end{eqnarray}
We could approximate the energy density of the synchrotron photon field $u_{\rm syn}(x)$ in the comoving frame by setting
\begin{eqnarray}
u_{\rm syn}(x)&=&\frac{1}{2\pi R(x)c}\int^{E_{s, max}}_{E_{s, min}}w_{\rm syn}(x,dx,E_{s})\nonumber\\&\times&e^{-\tau_{\rm tot}(E_{s},x)}dE_{s}\;.
\label{Eq:10}
\end{eqnarray}
\subsection{External ambient fields}
We consider the contribution of the external ambient field that includes both BLR and MT. We assume that: (i) the BLR is a shell located at a characteristic distance $R_{\rm BLR}\sim10^{17}(L_{d}/10^{45}\rm erg~s^{-1})^{1/2}$~cm (e.g. Ghisellini \& Tavecchio 2008), and its spectral shape observed in the comoving frame is a Maxwellian distribution peaking at photon energies of Lyman $\rm \alpha$ with $E_{\rm BLR}\sim 10 ~\rm eV\times \Gamma$; (ii) there is a MT (e.g. Blazejowski et al. 2000; Sikora et al. 2002) at characteristic distance $R_{\rm MT}\sim2.5\times10^{18}(L_{d}/10^{45}\rm erg~s^{-1})^{1/2}$~cm (e.g. Ghisellini \& Tavecchio 2008), and its spectral shape observed in the comoving frame is also a Maxwellian distribution peaking at photon energies of dust with $E_{\rm MT}\sim 0.3 ~\rm eV\times \Gamma$. Where $L_d$ is the luminosity of a accretion disc. We note that the defined initial location ($x=0$) of the $\gamma$-ray production site in the model would not show the real distance between the SMBH and the base of the jet. In order to derive a correlation between external ambient field and energy dissipation region, we use the distance $r$ between the position of dissipation region and the SMBH (e.g. Dermer et al. 2014) to parameterize the energy density of BLR and MT in the jet comoving frame (e.g. Hayashida et al. 2012; Zheng \& Yang 2016),
\begin{equation}
u_{\rm {BLR}}(r)=\frac{\tau_{\rm {BLR}}\Gamma^2 L_d}{3\pi R_{\rm{BLR}}^2c[1+(r/R_{\rm{BLR}})^{\beta_{\rm{BLR}}}]}\;,
\label{Eq:11}
\end{equation}
and
\begin{equation}
u_{\rm{MT}}(r)=\frac{\tau_{\rm{MT}}\Gamma^2 L_d}{3\pi R_{\rm{MT}}^2c[1+(r/R_{\rm{MT}})^{\beta_{\rm{MT}}}]}\;,
\label{Eq:12}
\end{equation}
where, $r=x_{0}+x$ with the real distance $x_{0}$ between the SMBH and the base of the jet in the comoving frame, $\tau_{\rm {BLR}}$ and $\tau_{\rm {MT}}$ are the fractions of the disc luminosity $L_d$ reprocessed into broad lines region and hot dust radiation, respectively. In our calculation, we adopt the radiation density profile $\beta_{\rm BLR}=3$ (e.g. Sikora et al. 2009) and $\beta_{\rm MT}=4$ (e.g. Hayashida et al. 2012), respectively. It can be seen that the BLR component is generated at distances $\sim$0.1 pc, much smaller than the MT that is produced at distances $\sim$1-10 pc. In both cases the photon densities decrease rather steeply with radius at distances beyond their characteristic radii, and achieve uniform densities at smaller distances.

\subsection{ICS off target photon fields}
We consider that target photons dominating ICS originate from both synchrotron photon and external ambient photon fields. In the model, we can write the total comoving target photon field $u_{\rm{tar}}(x)$ at the jet length $x$ through the expression
\begin{eqnarray}
u_{\rm{tar}}(x)&=&u_{\rm{syn}}(x)+u_{\rm {BLR}}(r=x_{0}+x)\nonumber\\&+&u_{\rm{MT}}(r=x_{0}+x)\;.
\label{Eq:13}
\end{eqnarray}
The Klein-Nishina (KN) effect is properly considered in the ICS by using fully KN cross section (e.g. Rybicki \& Lightman 1979; Blumenthal \& Gould 1979). We derived the total synchrotron emission and ICS contribution through integrating every segment from $x=0$ to $x=L$. Then we could reproduce the observed spectra at the Earth.

\section{Numerical Results}
\begin{figure}
	\centering
		\includegraphics[width=9.0 cm]{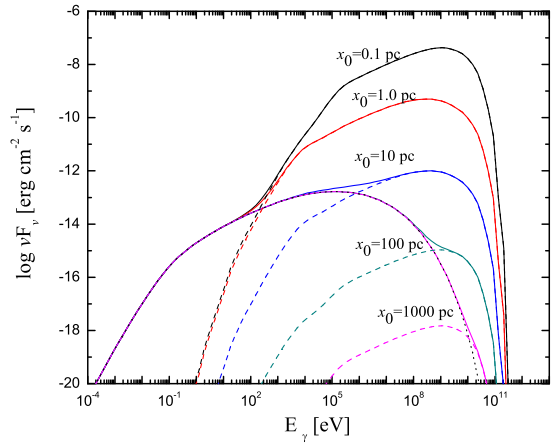}
		\caption{The expected ICs spectra with different initial energy dissipation region $x_{0}$. The dotted color curves represent ICs off the synchrotron photon fields, the dashed color curves represent ICs off the external ambient fields, and the solid color curves represent the total spectra. Marks near color curves represent the initial energy dissipation region $x_{0}$. We adopt the parameters as follows: $P_{j}=1.0\times10^{45}~\rm erg~s^{-1}$, $L=10$ Kpc, $B_{0}$=0.15 G, $A_{\rm equi}$=0.001, $E_{\rm min}$=5.0 MeV, $E_{\rm cut}=1.0\times10^{3}$ MeV, $L_{d}=6.0\times10^{45}~\rm erg~ s^{-1}$, $\tau_{\rm BLR}$=0.08, $\tau_{\rm MT}$=0.3, $\alpha$=2.0, $\theta_{\rm open}^{'}=2.0^{\circ}$, $\theta_{\rm obs}=3.0^{\circ}$, $\Gamma$=8.0.}
	\label{fig:1}
\end{figure}

We argue that, due to the dependence of the energy density of external ambient field on the energy dissipation region $x$, so the assumption of target photons dominating ICS from BLR, together with MT is valid. We test the expected ICS spectra with different initial energy dissipation region $x_{0}$ in Fig.{\ref{fig:1}}. We find that 1) the ICS spectra significantly depend on the location of initial energy dissipation region; 2) when the initial energy dissipation region $x_{0}$ is at large distances, we could neglect the contribution of the external ambient field. We emphasize though the parameter $x_{0}$ is a constant for a source, the energy density of external ambient field should change with the emission region moving along with the jet $x$-axis continuously.

\begin{figure*}[ht!]
  \begin{center}
   \begin{tabular}{cc}
        \includegraphics[scale=0.28]{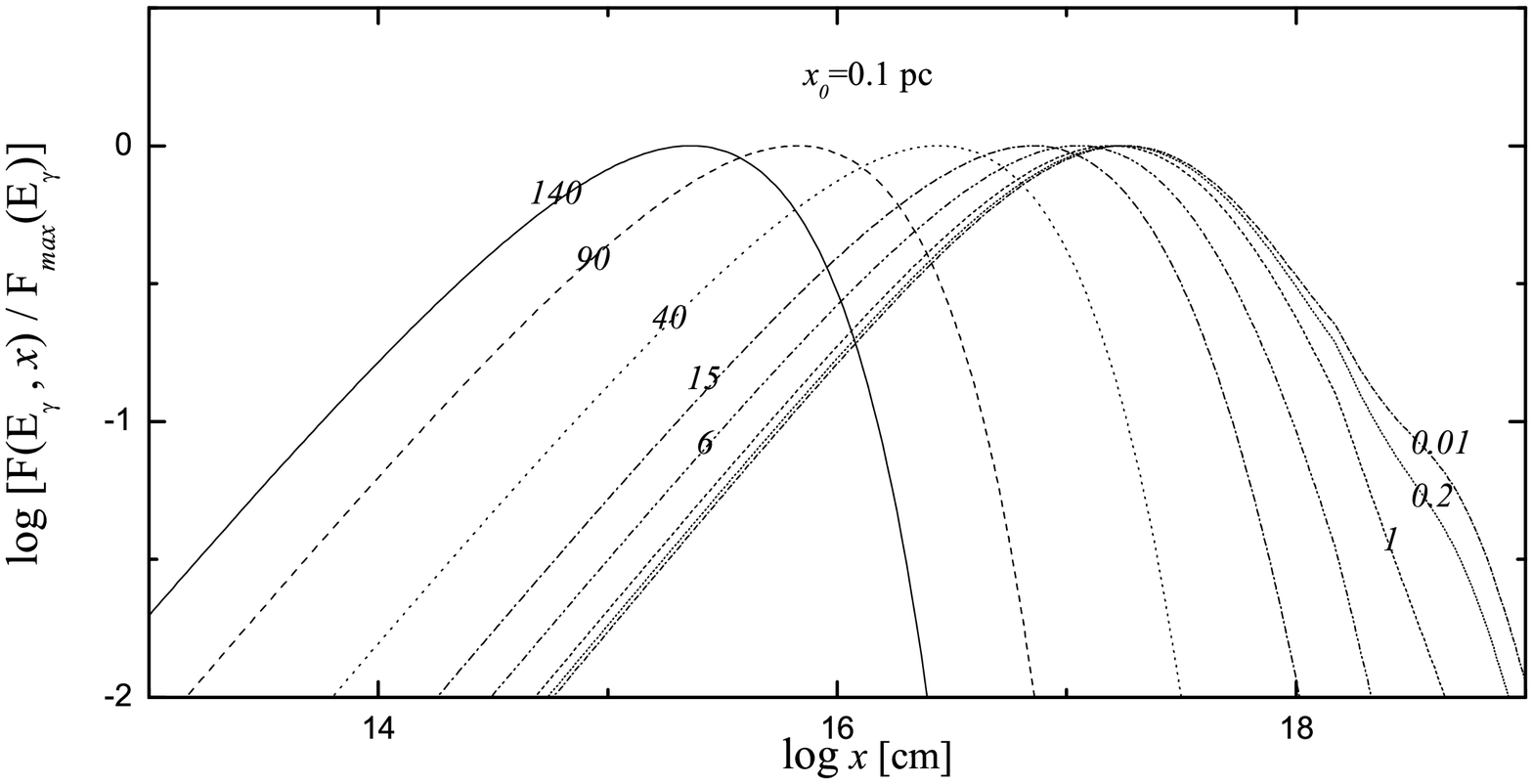}
        \includegraphics[scale=0.28]{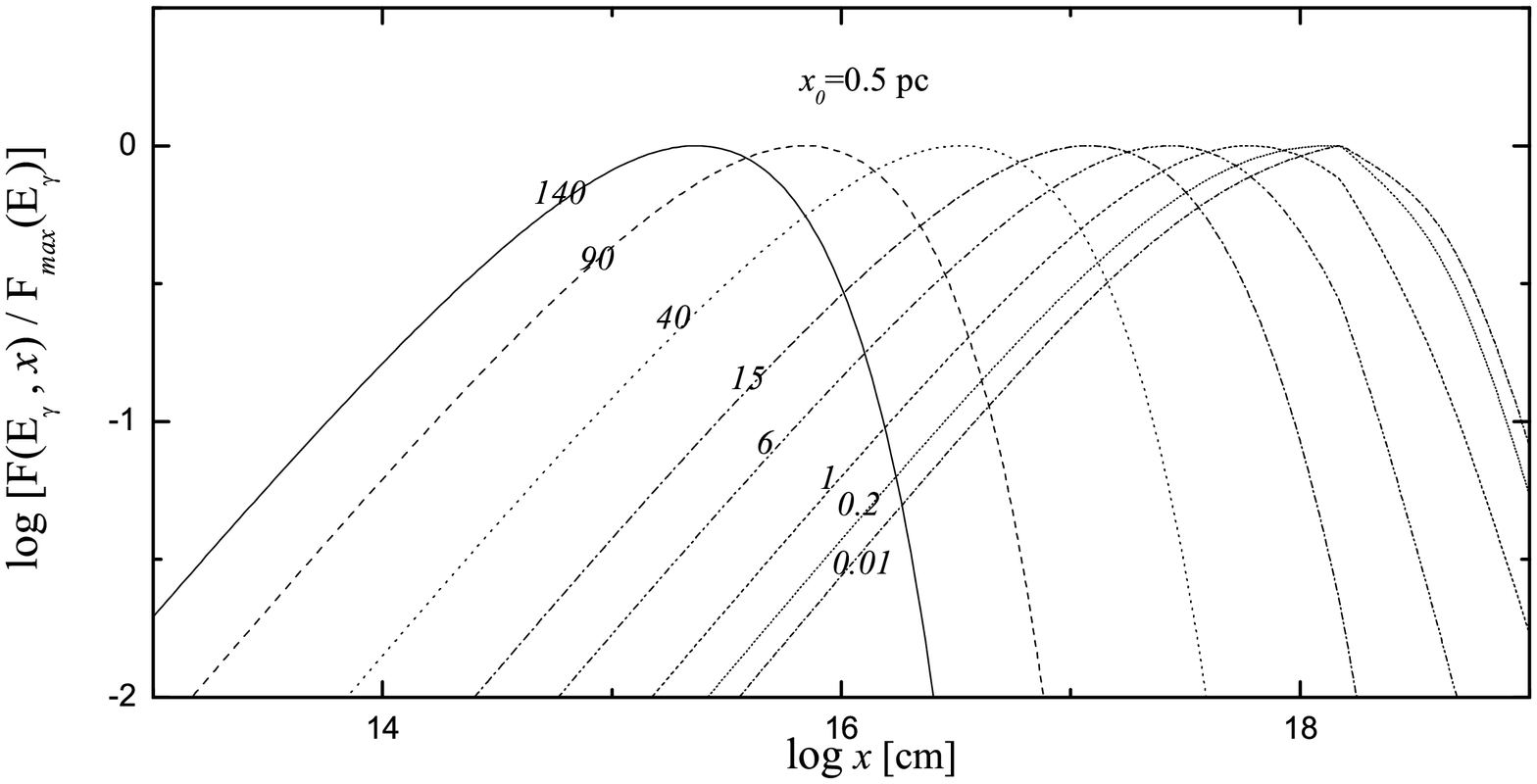}\\
        \includegraphics[scale=0.28]{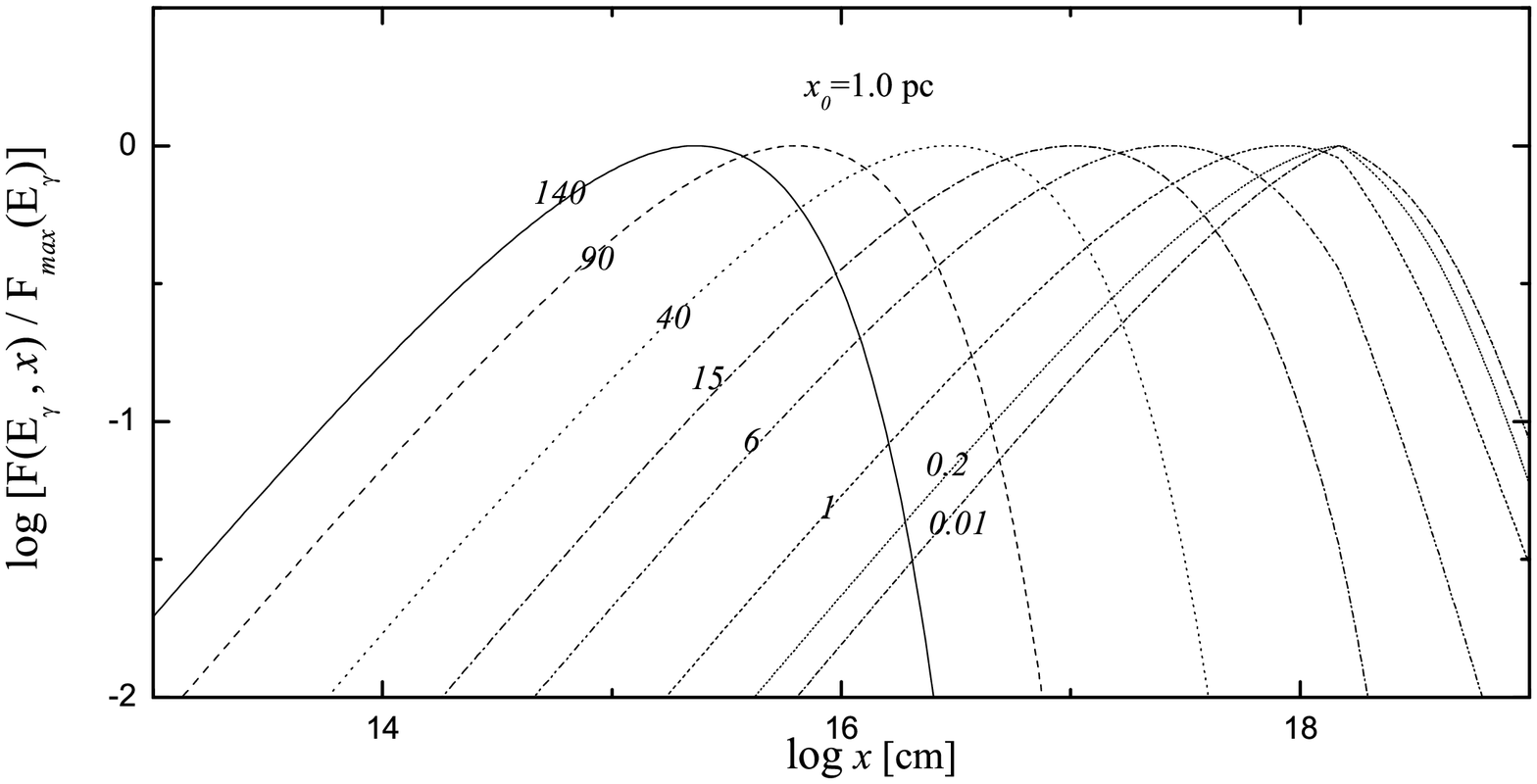}
        \includegraphics[scale=0.28]{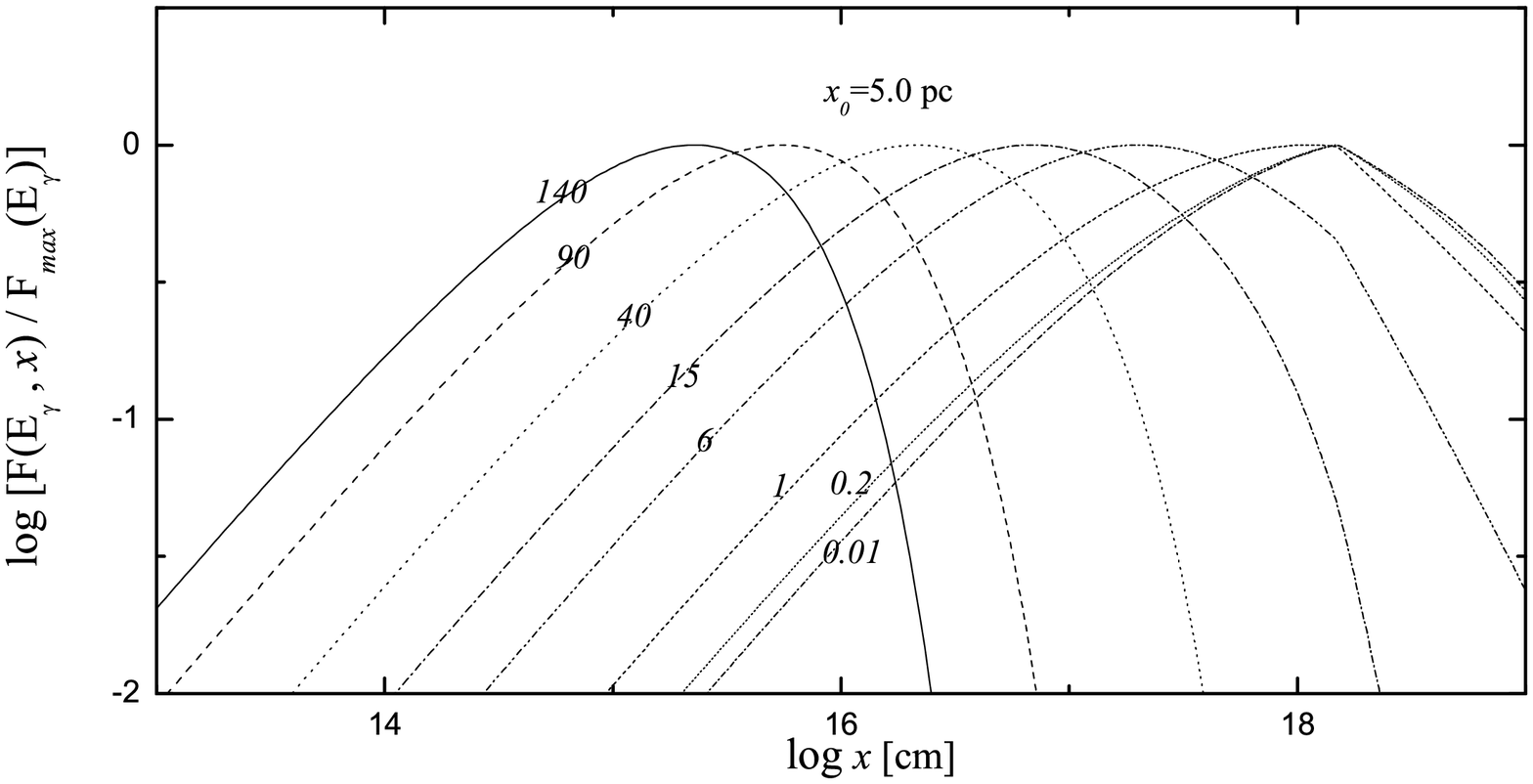}\\

    \end{tabular}
  \end{center}       		
	\caption{The length-dependent normalized ICs intensity in the different energy band. Marks near the curves represent the $\gamma$-ray photon energy in the unit of GeV. We adopt the parameters as follows: $P_{j}=1.0\times10^{45}~\rm erg~s^{-1}$, $L=10$ Kpc, $B_{0}$=0.15 G, $A_{\rm equi}$=0.001, $E_{\rm min}$=5.0 MeV, $E_{\rm cut}=1.0\times10^{3}$ MeV, $L_{d}=6.0\times10^{45}~\rm erg~s^{-1}$, $\tau_{\rm BLR}$=0.08, $\tau_{\rm MT}$=0.3, $\alpha$=2.0, $\theta_{\rm open}^{'}=2.0^{\circ}$, $\theta_{\rm obs}=3.0^{\circ}$, $\Gamma$=8.0.}
	\label{fig:2}
\end{figure*}

In order to penetrate the location of the $\gamma$-ray emission region, we show the length-dependent normalized ICs intensity in the different energy band in Fig.{\ref{fig:2}}. It can be seen that 1) the energetic $\gamma$-ray photons are produced in different emission regions; 2) the $\gamma$-ray photons with higher energy are produced near the base of the jet, on the contrary, the $\gamma$-ray photons with lower energy are produced at large distance from the base of the jet; 3) when the parameter $x_{0}$ increases, the MeV-GeV $\gamma$-ray emission region is far from the base of the jet with a constraint of $x\lesssim x_{0}$. As shows in Fig.{\ref{fig:2}}, there is very little change for such a large change in $x_{0}$. We note that the radius at distance of 5 pc are around the characteristic radius of MT, which shows that the photon densities do not obviously decrease.

In the present model, as a free parameter, the distance between the SMBH and the base of the jet $x_{0}$ determines the energy densities of BLR and MT at the base of the jet, and the intensity of the external Compton component. Because we derived the total ICS contribution through integrating every segment from $x=0$ to $x=L$, we change the energy dissipation region from $x_{0}$ to $x_{0}+L$ in the fluid frame. We propose that numerical results present a diagnostic on the location of the $\gamma$-ray emission region. Since the parameter $x_{0}$ could be obtained through finding a advisable external ambient field, we could determine the energy dissipation region by parameter $r=x_{0}+x$ through reproducing the $\gamma$-ray spectra.

\section{The Sample Description}
With the aim of tracing the the origin of external ambient photon fields in FSRQs jets, we expect the sample to satisfy the criterion: i) the source is observed intensively in MeV-GeV $\gamma$-ray energy band, and ii) a multi-wavelength simultaneous/quasi-simultaneous observations are available. Furthermore, we want to have the radio simultaneous/quasi-simultaneous data to constrain on the length and magnetism of the jet (Zheng \& Yang 2016). On the basis of above criterion,  we select 36 FSRQs from Giommi et al. (2012) and Paliya et al. (2013), where the multi-wavelength simultaneous/quasi-simultaneous observations by \emph{Plank}, \emph{Swift}, \emph{Fermi}-LAT and other some ground-based telescopes are accumulated. The source names of the sample are listed in Table {\ref{Table:1}}. Because the simultaneous observational data in the radio,  optical and X-ray energy bands are not available for the sources PKS 0215+015, PKS 0528+134 and PKS 1502+036, we adopt the literature or archival data in the radio, optical, and X-ray energy bands. Since accurate SEDs in MeV-GeV $\gamma$-ray energy band are expected, we also contain the \emph{Fermi}-LAT data integrated over a period of 27 months from August 4, 2008 to November 4, 2010.

\section{Applications}
\subsection{Modelling the SEDs}
Using the synchrotron radiation and ICS solution for the conical jet structure, we can reproduce the SED for each source in the sample. We note that the spectrum derived {\bf from} the model is described by the geometry and physical parameters of the jet ($P_{j}$, $L$, $B_{0}$, $x_{0}$, $A_{\rm equi}$, $\theta_{\mathrm{open}}$, $\theta_{\mathrm{obs}}$ and $\Gamma$), the initial injected electron distribution ($\alpha$, $E_{\rm min}$ and $E_{\rm cut}$) and the external ambient fields ($L_{d}$, $\tau_{\rm BLR}$ and $\tau_{\rm MT}$). It is well known that these parameters are not directly observable. On the other hand, they are coupled with each other in the framework of the length-dependent conical jet model. In these scenarios, it will take too long time to reproduce the best SED if we allow all of the parameters to be free. In order to determine the parameters reliable, we list some constraints that characterize the parameters:

(1) statistical results show that the typical jet powers of FSRQs are in the range of $10^{45}~\rm erg~s^{-1}\leq P_{j}\leq10^{48}~\rm erg~s^{-1}$ (e.g. Ghisellini \& Tavecchio 2008; Ghisellini 2010b; Kang et al. 2015);

(2) we could give a constraint on the length of the jet $L$ and the magnetic field $B_{0}$ at the base of the jet from a radio spectrum (Zheng \& Yang 2016);

(3) the angle of observer's line of sight to the jet axis in the lab frame is $\theta_{\mathrm{obs}}\leq5^\circ$ (Urry 1995);

(4) electron spectra index $\alpha$ is thought to be between 1 and 3 from the theory of the shock acceleration (Bell 1978; Bell et al. 2011; Summerlin \& Baring 2012);

(5) we adopt a simplified Shakura-Sunyaev disk spectrum (Dermer et al. 2014) as follows:
\begin{equation}
\epsilon L_{d}(\epsilon)=1.12 L_{d}(\frac{\epsilon}{\epsilon_{\rm max}})^{4/3}\exp(-\frac{\epsilon}{\epsilon_{\rm max}})
\end{equation}
where $\epsilon$ is the photon energy that is emitted by the accretion disk. While the value of $\epsilon_{\rm max}$ depends on the spin of the black hole and relative Eddington luminosity, we set a typical characteristic temperature of the UV bump in Seyfert galaxies with $\epsilon_{\rm max}\sim 10$ eV. The accretion disk produces a total luminosity $L_{d}=\eta\dot{M}c^{2}$, where $\dot{M}$ is the accretion rate and $\eta$ is the accretion efficiency. It is known that the efficiency of conversion for nuclear reaction is $\eta=0.007$ and that for pure accretion is $\eta=0.1$. If we set $\eta=0.08$ (e.g. Ghisellini 2010b), and we adopt a Eddington accretion rate $\dot{M}=1.39\times10^{18}(M/M_{\odot})~\rm g~s^{-1}$ with $M\sim10^{9}M_{\odot}$, we could estimate an accretion-disk luminosity $L_{d}\lesssim 10^{46} \rm erg~s^{-1}$.

Following the above constraints, we could search for the electron and photon spectra along $x$-axis of the jet. Assuming a steady geometry of the jet structure for a source, and a immovable position of the observer during in observed epoches, we calculate the electron and photon spectrum in every segment from $x=0$ to $x=L$. We consider that observed spectrum is a summation of each segment. Therefore, we can calculate the observed spectrum using the photon spectrum in every segments from $x=0$ to $x=L$. We list these parameters in Table {\ref{Table:1}}. The columns in this table are as follows.

1. source name;

2. $z$, redshift;

3. $P_{j}$, total jet power in the unit of $\rm erg~s^{-1}$;

4. $L$, the length of the jet in the unit of pc;

5. $B_0$, the magnetic field at the base of jet in the unit of G;

6. $x_{0}$, the distance between the SMBH and base of jet in the unit of pc;

7. $E_{\rm e,min}$, the minimum energy of initial injected electron in the unit of MeV;

8. $E_{\rm e,cut}$, the cut off energy of initial injected electron in the unit of MeV;

9. $L_d$, the luminosity of a accretion disc in the unit of $\rm erg~s^{-1}$;

10. $\tau_{\rm {BLR}}$, the fraction of the disc luminosity reprocessed into broad lines region;

11. $\tau_{\rm {MT}}$, the fraction of the disc luminosity $L_d$ reprocessed into hot dust radiation;

12. $\alpha$, the electron spectra index;

13. $\theta^\prime_{\rm open}$, the half opening angle of the cone;

14. $\theta_{\mathrm{obs}}$, the angle of observer's line of sight to the jet axis;

15. $\Gamma$, the bulk Lorentz factor;

16. $N$, the number of simultaneous observational data points;

17. $\chi^{2}$, the $\chi^{2}=\frac{1}{N-dof}\sum\limits_{i=1}^{N}(\frac{\Hat{y_{i}}-y_{i}}{\sigma_{i}})^{2}$, where $dof$ are the degrees of freedom, i.e. the number of free parameters used for the model. The $\Hat{y_{i}}$ are the expected values from the model and the $y_{i}$ are the observed data. $\sigma_{i}$ is the standard deviation for each data point. We take 1\% of the observed radio and optical flux and take 2\% of the observed UV and x-ray flux
as the errors of the data points whose errors are available (e.g., Zhang et al. 2012; Aleksi¡äc et al. 2014).


\begin{table*}[htbp]
\centering
\tiny
\caption{The parameters of model spectra \label{Table:1}.}
\tabcolsep 1.20mm
\begin{tabular}{llcl llllc cllll llllc ll}
\hline \hline
Source name & $z$ & $ P_{\rm j}$ & $L$  & $B_{\rm 0}$ & $x_{0}$ & $A_{equi.}$ & $E_{min}$ & $E_{\rm cut}$ & $L_{d}$ & $\tau_{BLR}$ & $\tau_{MT}$ & $\alpha$ & $\theta_{open.}^{'}$ & $\theta_{obs.}$  &  $\Gamma$ & $N$ & $\chi^{2}$ \\
~... & ... & $erg~s^{-1}$ & pc & G &	pc & ... &	MeV & MeV & $erg~s^{-1}$ & ... & ... & ... & ... & ... & ... & ... & ... \\
\hline

\uppercase\expandafter{\romannumeral3} ZW 2	&	0.089 	&	$2.00\times45$	&	32	&	0.95 	&	14.47 	&	0.001	 &	5.11	&	$6.20\times3$	&	$2.30\times45$	&	0.10 	&	0.30 	&	2.1	&	3.0$^\circ$ 	&	 2.0$^\circ$ 	&	6.0 	&	28 	&	1.50	\\
S4 0133+47	&	0.859 	&	$7.55\times46$	&	40	&	0.10 	&	1.29 	&	0.045	&	10.11	&	$2.11\times3$	 &	$6.11\times45$	&	0.10 	&	0.30 	&	2.1	&	3.5$^\circ$ 	&	2.3$^\circ$ 	&	8.5 	&	29 	&	 47.24	\\
PKS 0202-17	&	1.740 	&	$1.90\times47$	&	94	&	0.15 	&	5.97 	&	0.015	&	5.11	&	$9.00\times3$	 &	$6.51\times45$	&	0.80 	&	0.85 	&	2.1	&	3.0$^\circ$ 	&	2.0$^\circ$ 	&	12.0 	&	34 	&	 3.30	\\
PKS 0215+015	&	1.715 	&	$3.10\times48$	&	94	&	0.40 	&	3.14 	&	0.1	&	11.11	&	$8.00\times3$	 &	$2.20\times45$	&	0.10 	&	0.30 	&	2.6	&	3.0$^\circ$ 	&	2.0$^\circ$ 	&	50.0 	&	34 	&	 11.12	\\
4C 28.7	        &	1.213 	&	$5.85\times46$	&	950	&	0.45 	&	1.70 	&	0.09	&	6.11	&	 $1.01\times3$	&	$3.12\times45$	&	0.10 	&	0.50 	&	2.1	&	3.0$^\circ$ 	&	2.0$^\circ$ 	&	14.0 	 &	29 	&	16.98	\\
PKS 0420-01	&	0.916 	&	$1.50\times47$	&	97	&	0.15 	&	20.44 	&	0.015	&	11.61	&	$1.82\times3$	 &	$3.50\times45$	&	0.10 	&	0.30 	&	2.1	&	3.0$^\circ$ 	&	2.0$^\circ$ 	&	16.0 	&	51 	&	 4.74	\\
PKS 0454-234	&	1.003 	&	$3.02\times46$	&	32	&	0.30 	&	0.41 	&	0.15	&	5.11	&	 $1.75\times3$	&	$5.11\times45$	&	0.10 	&	0.30 	&	2.1	&	3.0$^\circ$ 	&	2.0$^\circ$ 	&	10.0 	 &	25 	&	29.79	\\
PKS 0528+134	&	2.070 	&	$2.50\times47$	&	975	&	0.15 	&	0.82 	&	0.015	&	5.11	&	 $1.71\times3$	&	$4.61\times45$	&	0.40 	&	0.50 	&	2.1	&	3.0$^\circ$ 	&	2.0$^\circ$ 	&	10.0 	 &	43	&	2.29	\\
1Jy 0537-286	&	3.104 	&	$4.70\times47$	&	102	&	0.20 	&	1.73 	&	0.04	&	5.14	&	 $1.10\times3$	&	$6.30\times45$	&	0.30 	&	0.60 	&	2.1	&	3.0$^\circ$ 	&	2.0$^\circ$ 	&	15.0 	 &	16 	&	30.23	\\
4C 71.07	&	2.218 	&	$4.31\times47$	&	189	&	0.22 	&	2.64 	&	0.016	&	6.11	&	$2.10\times3$	 &	$6.51\times45$	&	0.83 	&	0.86 	&	2.1	&	3.0$^\circ$ 	&	2.0$^\circ$ 	&	12.0 	&	53 	&	 3.76	\\
S4 0917+44	&	2.190 	&	$8.70\times47$	&	197	&	0.32 	&	2.55 	&	0.1	&	5.11	&	$1.10\times3$	&	 $5.80\times45$	&	0.01 	&	0.30 	&	2.1	&	3.0$^\circ$ 	&	2.0$^\circ$ 	&	28.0 	&	53 	&	0.92	 \\
4C 55.17	        &	0.896 	&	$1.50\times46$	&	35	&	0.51 	&	1.64 	&	0.15	&	5.11	&	 $3.10\times3$	&	$5.81\times45$	&	0.40 	&	0.40 	&	2.1	&	3.0$^\circ$ 	&	2.0$^\circ$ 	&	10.0 	 &	22 	&	14.46	\\
PKS 1124-186	&	1.048 	&	$2.30\times46$	&	67	&	0.70 	&	0.28 	&	0.65	&	5.11	&	 $2.01\times3$	 &	$4.21\times45$	&	0.05 	&	0.30 	&	2.1	&	3.0$^\circ$ 	&	2.0$^\circ$ 	&	9.0 	 &	44 	&	 6.96	\\
PKS 1127-145	&	1.184 	&	$3.21\times46$	&	109	&	0.80 	&	1.26 	&	0.1	&	5.11	&	$1.33\times3$	 &	$2.01\times45$	&	0.01 	&	0.88 	&	2.1	&	1.2$^\circ$ 	&	1.0$^\circ$ 	&	10.5 	&	40 	&	 204.07	\\
4C 49.22	        &	0.334 	&	$7.21\times45$	&	94	&	0.80 	&	4.43 	&	0.015	&	5.11	&	 $1.01\times3$	&	$3.05\times45$	&	0.01 	&	0.80 	&	2.1	&	3.0$^\circ$ 	&	2.5$^\circ$ 	&	9.0 	 &	58 	&	3.46	\\
4C 29.45	&	0.725 	&	$5.21\times45$	&	23	&	0.40 	&	2.24 	&	0.01	&	5.9	&	$1.01\times3$	&	 $1.11\times45$	&	0.10 	&	0.30 	&	2.1	&	0.2$^\circ$ 	&	0.1$^\circ$ 	&	15.0 	&	41 	&	54.25	 \\
PKS 1219+04	&	0.965 	&	$4.01\times46$	&	96	&	0.50 	&	1.64 	&	0.015	&	7.11	&	$1.51\times3$	 &	$1.01\times45$	&	0.08 	&	0.96 	&	2.1	&	3.0$^\circ$ 	&	2.0$^\circ$ 	&	12.0 	&	27 	&	 18.01	\\
3C 273	        &	0.158 	&	$1.91\times46$	&	158	&	0.90 	&	1.10 	&	0.015	&	5.11	&	 $1.01\times3$	&	$2.01\times45$	&	0.10 	&	0.30 	&	2.1	&	3.0$^\circ$ 	&	2.0$^\circ$ 	&	6.0 	 &	46 	&	126.45	\\
PKS 1244-255	&	0.635 	&	$1.70\times46$	&	95	&	0.60 	&	0.95 	&	0.015	&	5.11	&	 $7.01\times3$	 &	$1.81\times45$	&	0.10 	&	0.30 	&	2.1	&	3.0$^\circ$ 	&	2.0$^\circ$ 	&	8.0 	 &	48 	&	 5.24	\\
3C 279	        &	0.533 	&	$3.50\times46$	&	96	&	0.40 	&	0.60 	&	0.15	&	5.11	&	 $1.02\times3$	&	$6.01\times45$	&	0.10 	&	0.30 	&	2.1	&	3.0$^\circ$ 	&	2.0$^\circ$ 	&	7.0 	 &	57 	&	41.74	\\
PKS 1502+106	&	1.839 	&	$3.32\times47$	&	4717	&	0.29 	&	1.95 	&	0.15	&	5.11	&	 $1.20\times3$	&	$9.00\times45$	&	0.10 	&	0.30 	&	2.4	&	3.0$^\circ$ 	&	2.0$^\circ$ 	&	25.0 	 &	50 	&	5.12	\\
PKS 1502+36	&	0.408 	&	$8.50\times45$	&	95	&	0.75 	&	3.18 	&	0.06	&	4.11	&	$1.01\times3$	 &	$1.01\times45$	&	0.05 	&	0.40 	&	2.5	&	3.0$^\circ$ 	&	2.0$^\circ$ 	&	18.0 	&	28	&	 17.13	\\
4C 38.41	        &	1.814 	&	$1.20\times47$	&	97	&	0.15 	&	1.82 	&	0.06	&	5.11	&	 $1.10\times3$	&	$4.30\times45$	&	0.10 	&	0.97 	&	2.1	&	3.0$^\circ$ 	&	2.0$^\circ$ 	&	13.0 	 &	51 	&	4.02	\\
3C 345	        &	0.593 	&	$4.80\times46$	&	94	&	0.40 	&	1.32 	&	0.15	&	1.11	&	 $3.00\times3$	&	$3.00\times45$	&	0.10 	&	0.98 	&	2.6	&	3.0$^\circ$ 	&	2.0$^\circ$ 	&	11.0 	 &	49 	&	39.01	\\
S5 1803+784	&	0.680 	&	$6.00\times46$	&	41	&	0.90 	&	4.09 	&	0.05	&	3.11	&	$6.00\times3$	 &	$5.33\times45$	&	0.10 	&	0.30 	&	2.6	&	3.0$^\circ$ 	&	2.0$^\circ$ 	&	18.0 	&	41 	&	 15.70	\\
PKSB 1908-201	&	1.119 	&	$3.01\times46$	&	93	&	0.60 	&	1.20 	&	0.4	&	5.11	&	$2.99\times3$	 &	$7.31\times45$	&	0.87 	&	0.88 	&	2.1	&	2.5$^\circ$ 	&	2.0$^\circ$ 	&	8.0 	&	32 	&	 21.17	\\
PMNJ 1923-2104	&	0.874 	&	$1.71\times46$	&	32	&	0.76 	&	2.83 	&	0.1	&	30	&	$1.01\times3$	&	 $1.61\times45$	&	0.10 	&	0.30 	&	2.1	&	3.0$^\circ$ 	&	0.5$^\circ$ 	&	18.0 	&	32 	&	19.44	 \\
OV-236	        &	0.352 	&	$3.00\times46$	&	95	&	0.15 	&	2.20 	&	0.015	&	8.11	&	 $1.51\times3$	&	$2.13\times45$	&	0.10 	&	0.30 	&	2.1	&	3.0$^\circ$ 	&	2.0$^\circ$ 	&	8.0 	 &	39 	&	3.15	\\
4C 06.69	        &	0.990 	&	$9.60\times46$	&	97	&	0.12 	&	8.81 	&	0.009	&	8.11	&	 $6.20\times3$	&	$1.10\times45$	&	0.10 	&	0.30 	&	2.1	&	3.0$^\circ$ 	&	1.0$^\circ$ 	&	8.0 	 &	45 	&	22.89	\\
PKS 2149-307	&	2.345 	&	$4.80\times47$	&	975	&	0.22 	&	4.09 	&	0.002	&	6.11	&	 $1.21\times3$	&	$2.60\times45$	&	0.10 	&	0.72 	&	2.1	&	3.0$^\circ$ 	&	0.8$^\circ$ 	&	19.0 	 &	30 	&	18.42	\\
4C 31.63	&	0.295 	&	$7.91\times45$	&	97	&	0.75 	&	4.09 	&	0.015	&	5.11	&	$1.01\times3$	 &	$2.31\times45$	&	0.01 	&	0.70 	&	2.1	&	3.0$^\circ$ 	&	2.0$^\circ$ 	&	8.0 	&	30 	&	 69.04	\\
PKS 2204-54	&	1.206 	&	$4.50\times46$	&	975	&	0.70 	&	1.92 	&	0.04	&	5.11	&	$2.60\times3$	 &	$1.20\times45$	&	0.09 	&	0.80 	&	2.3	&	3.0$^\circ$ 	&	2.0$^\circ$ 	&	13.0 	&	31 	&	 17.58	\\
PKS 2227-08	&	1.560 	&	$2.20\times47$	&	97	&	0.15 	&	1.76 	&	0.015	&	5.11	&	$2.41\times3$	 &	$4.90\times45$	&	0.08 	&	0.90 	&	2.3	&	3.0$^\circ$ 	&	2.0$^\circ$ 	&	11.0 	&	47 	&	 62.76	\\
4C 11.69	&	1.037 	&	$7.01\times46$	&	224	&	0.15 	&	1.98 	&	0.06	&	7.11	&	$3.41\times3$	 &	$5.41\times45$	&	0.08 	&	0.30 	&	2.1	&	3.0$^\circ$ 	&	2.0$^\circ$ 	&	8.0 	&	41 	&	 91.17	\\
3C 454.3	&	0.859 	&	$3.20\times47$	&	97	&	0.60 	&	0.60 	&	0.05	&	5.11	&	$5.10\times3$	 &	$5.20\times45$	&	0.10 	&	0.30 	&	2.1	&	3.0$^\circ$ 	&	2.0$^\circ$ 	&	12.0 	&	57 	&	 5.09	\\
PKS 2325+093	&	1.843 	&	$2.50\times47$	&	97	&	0.50 	&	3.02 	&	0.02	&	5.11	&	 $1.10\times3$	&	$5.50\times45$	&	0.10 	&	0.30 	&	2.3	&	3.0$^\circ$ 	&	2.0$^\circ$ 	&	18.0 	 &	23 	&	19.44	\\

\hline
\end{tabular}
\end{table*}

In Figures 3-8, we show the multi-wavelength spectra of the sources in the sample, respectively. For comparison, the observed data of the sources are also shown. In these figures, the simultaneous data are shown in red; The quasi-simultaneous data including \emph{Fermi}-LAT data over 2 months, \emph{Planck} ERCSC and non-simultaneous ground based observations are shown in green; The \emph{Fermi}-LAT data integrated over 27 months are shown in blue; the literature or archival data are shown in gray. The dashed line represents the synchrotron emission, the dotted line represents ICS on the seed photons of the synchrotron, BLR and MT, and the thick solid line represents the total spectrum by summation all of emission components respectively. It can be seen that the observed data can be reproduced in the model.

\begin{figure*}[ht!]
  \begin{center}
   \begin{tabular}{cc}
		\includegraphics[scale=0.2]{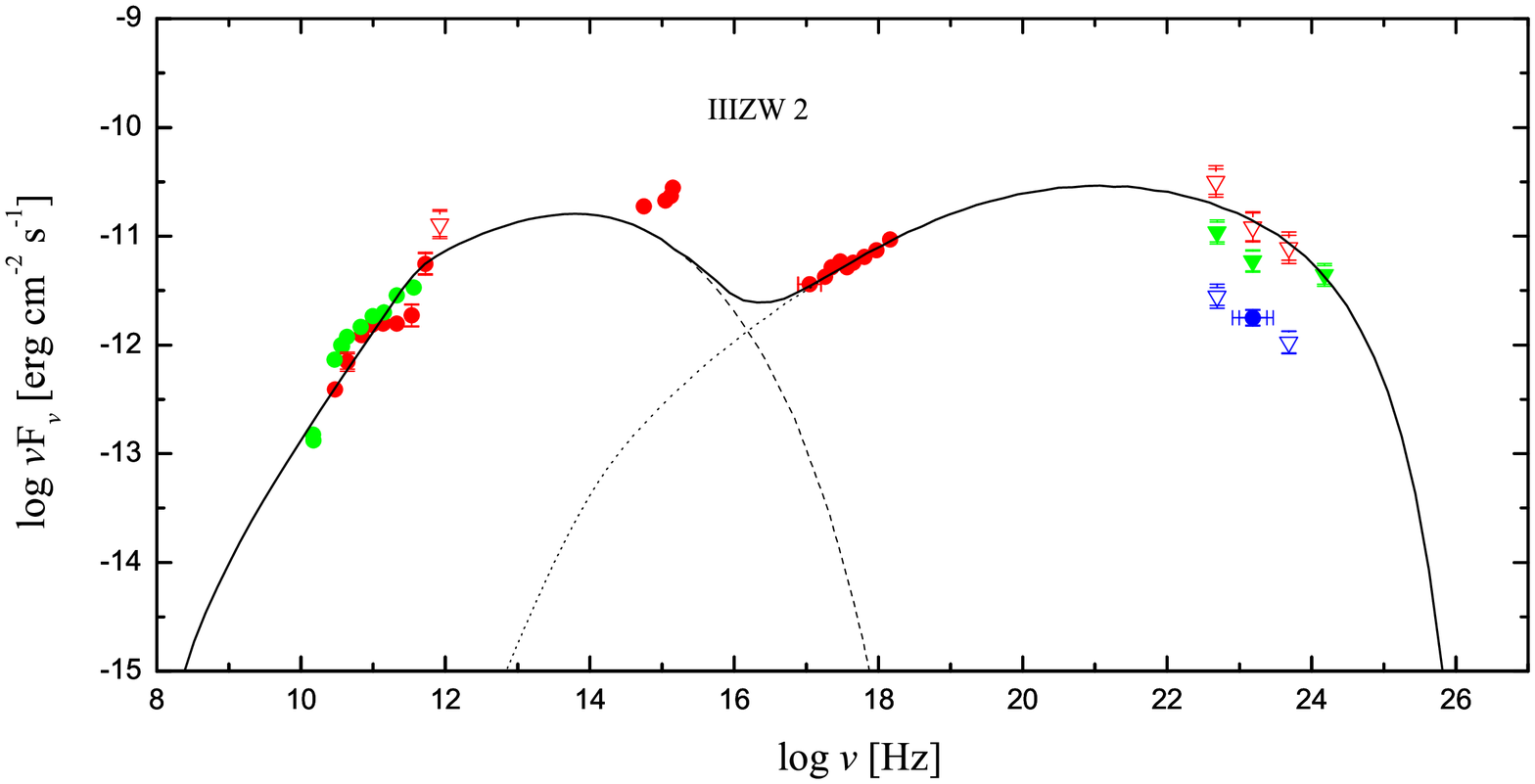}
        \includegraphics[scale=0.2]{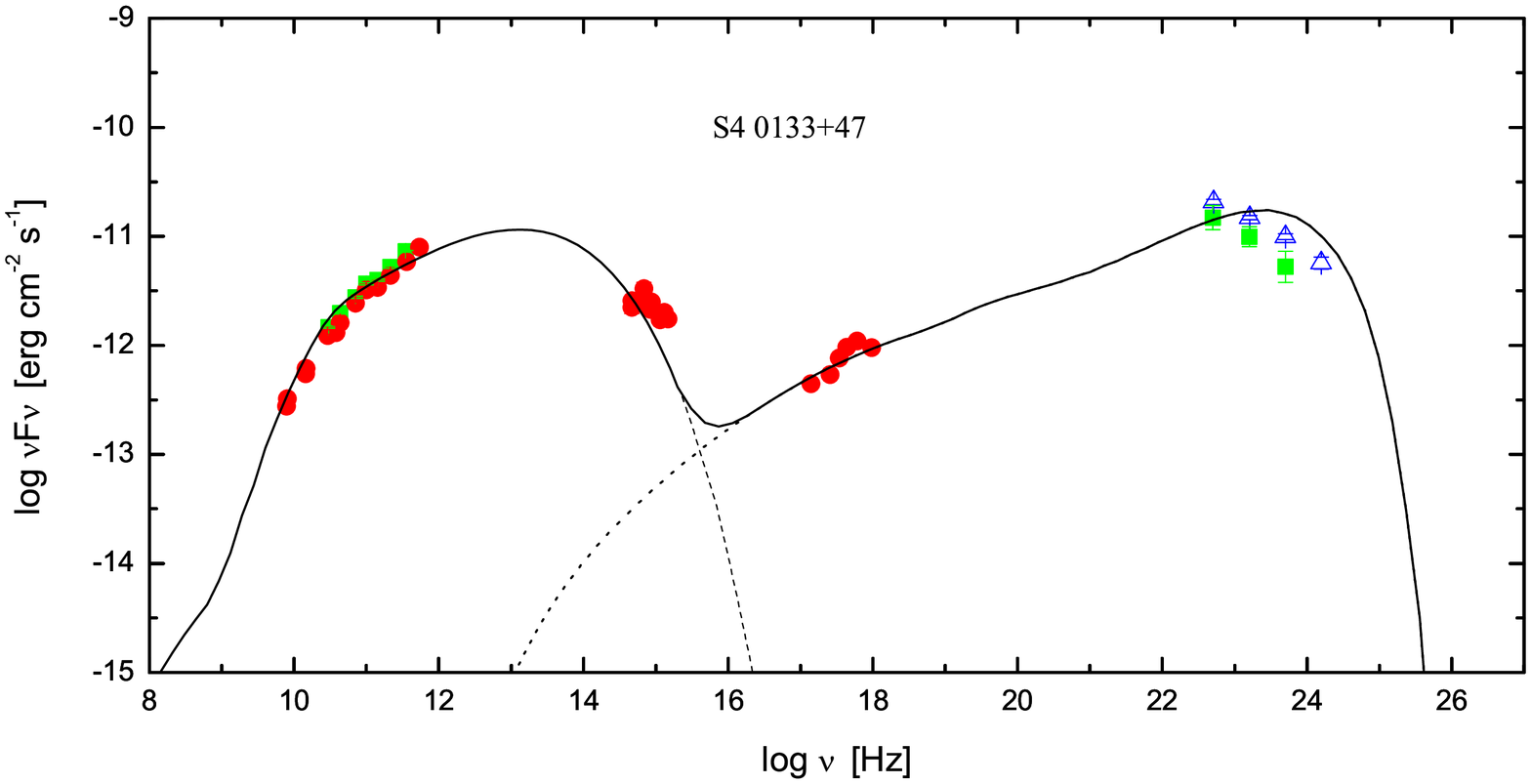}\\
        \includegraphics[scale=0.2]{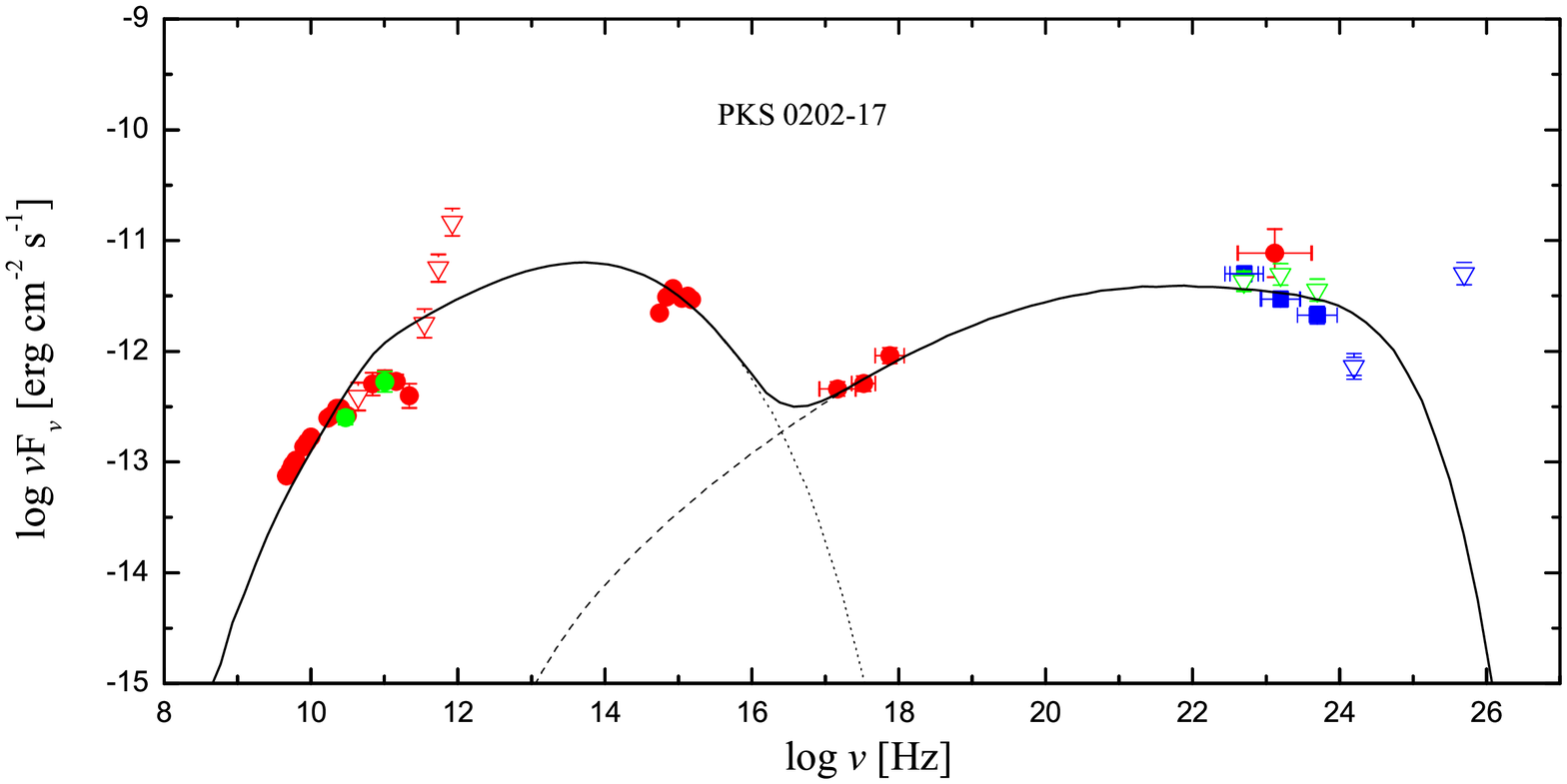}
        \includegraphics[scale=0.2]{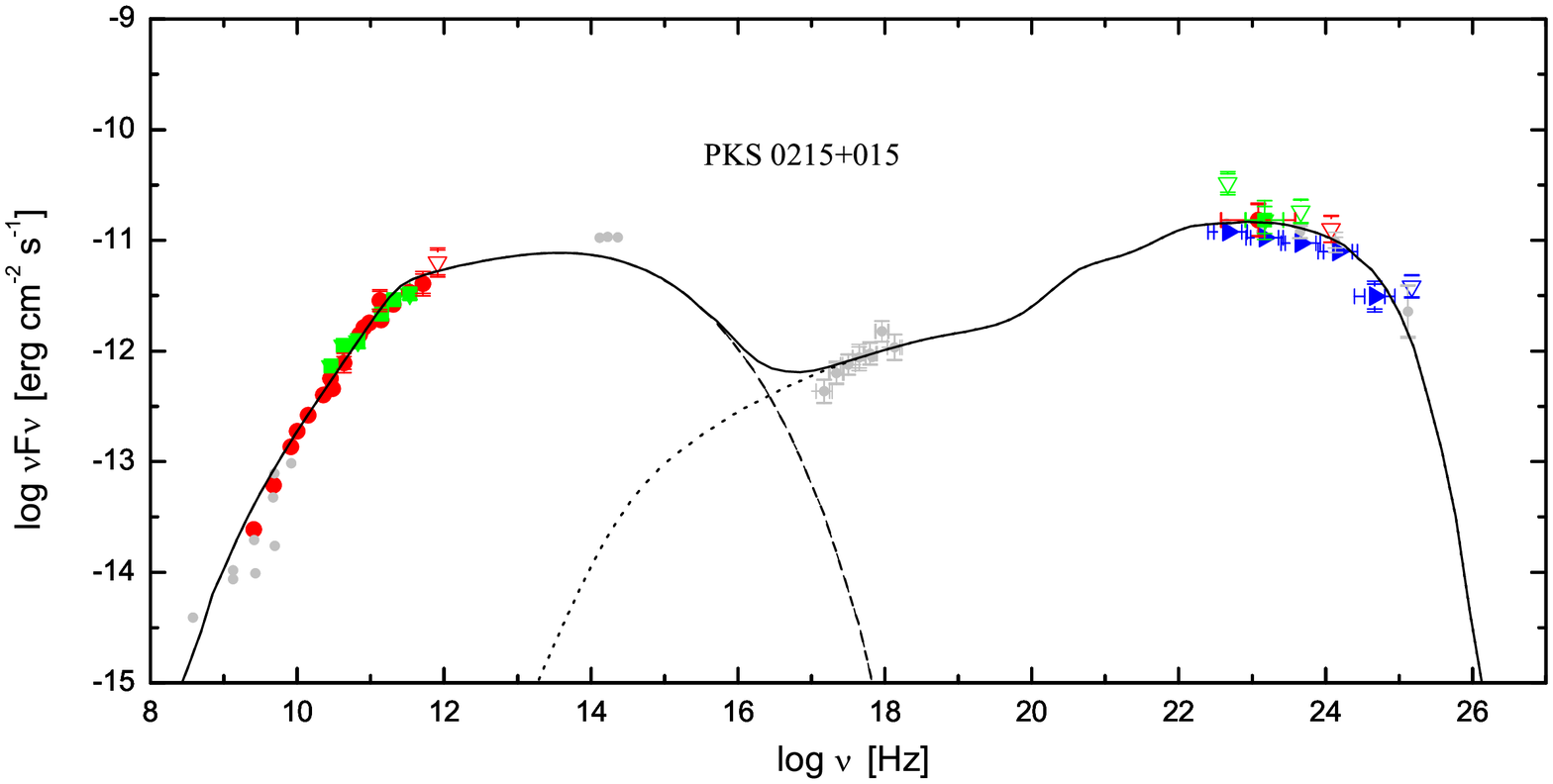}\\
        \includegraphics[scale=0.2]{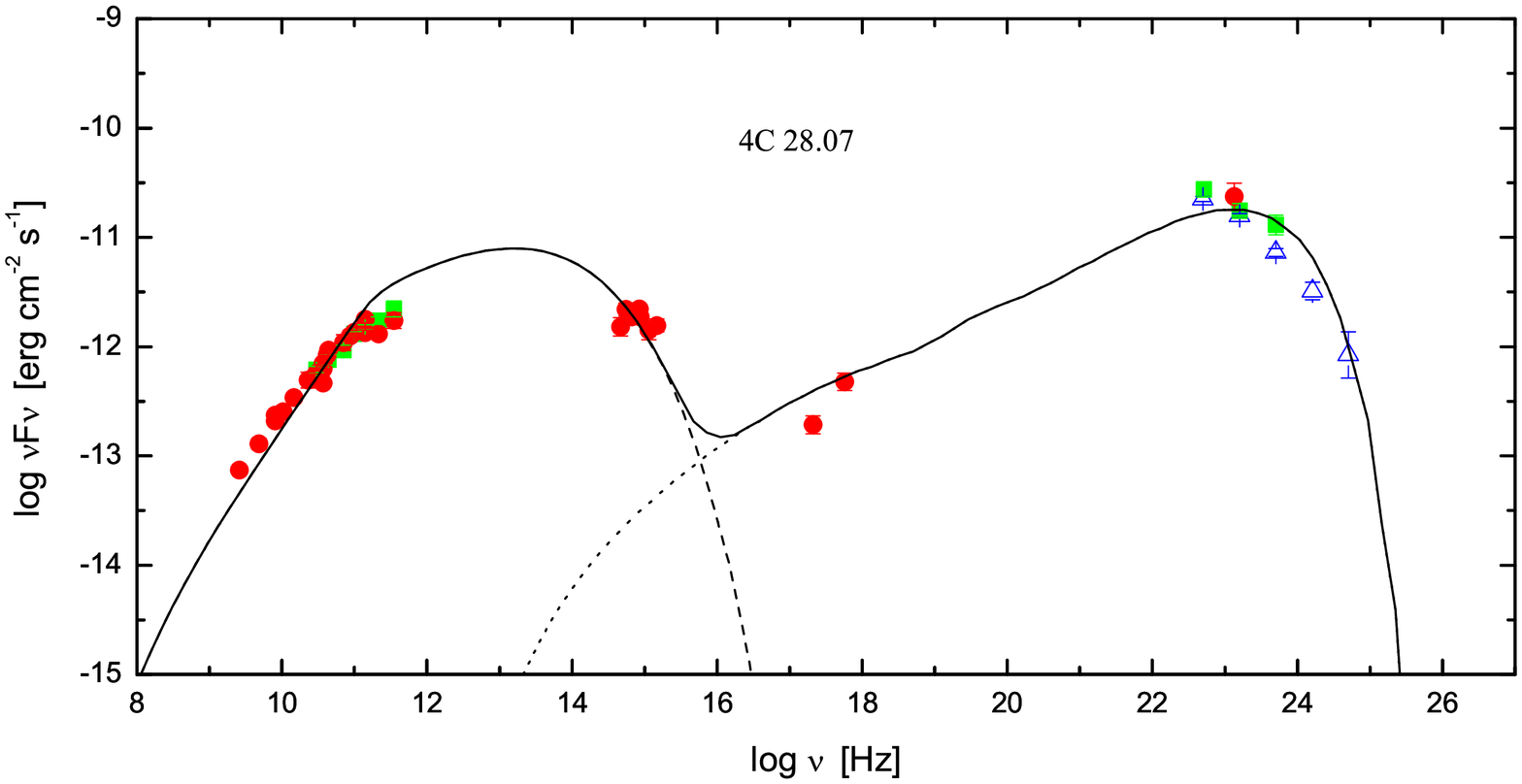}
        \includegraphics[scale=0.2]{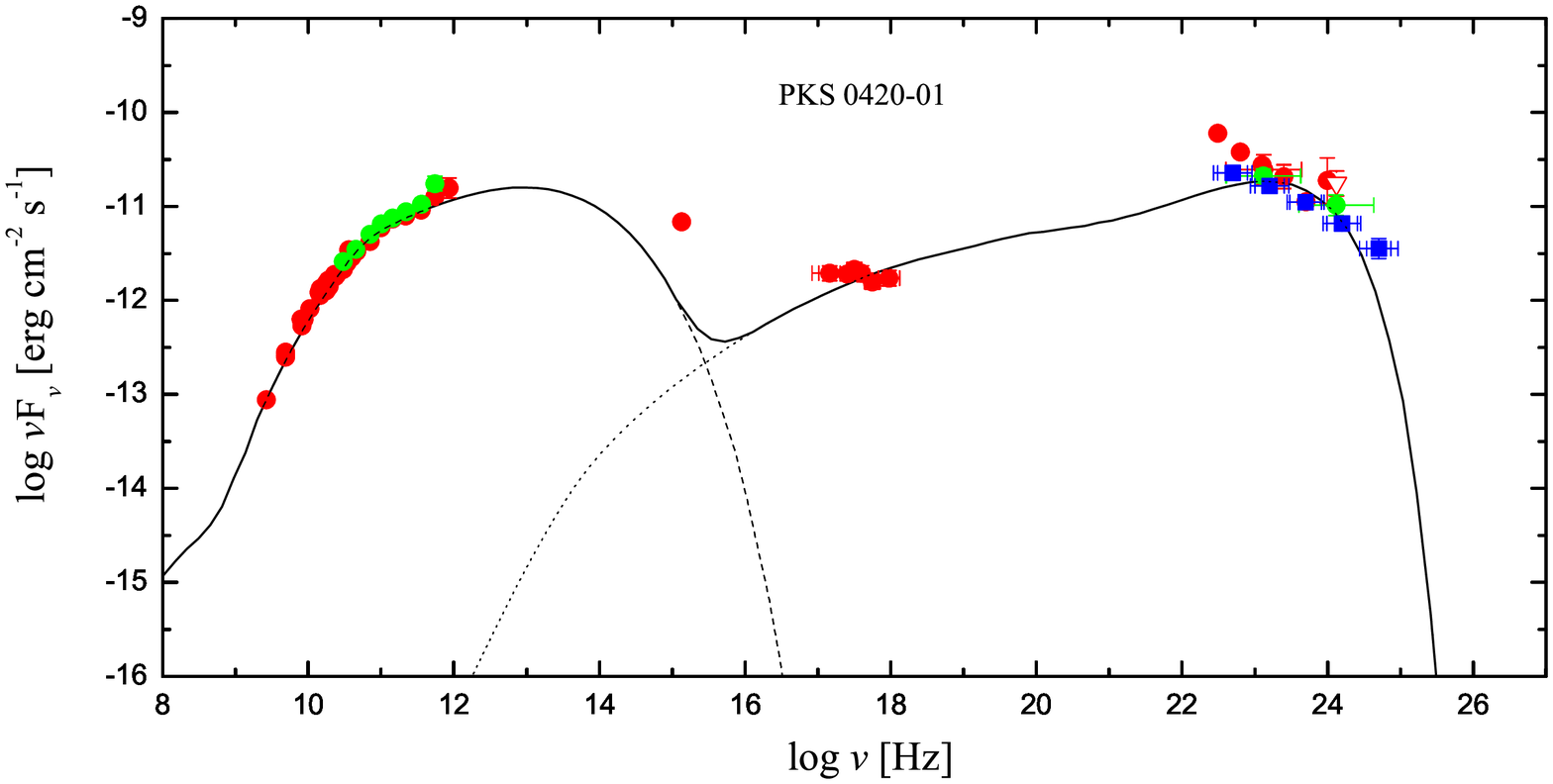}\\
   \end{tabular}
  \end{center}  		
	\caption{Comparisons of predicted multi-wavelength spectra with observed data for \uppercase\expandafter{\romannumeral3}ZW 2, PKS 0202-17, PKS 0215+015, S4 0133+47, 4C 28.07, PKS 0420-01, respcetively. The simultaneous data are shown in red; The quasi-simultaneous data including \emph{Fermi}-LAT data over 2 months, \emph{Planck} ERCSC and non-simultaneous ground based observations are shown in green; The \emph{Fermi}-LAT data integrated over 27 months are shown in blue; the literature or archival data are shown in gray. The dashed line represents the synchrotron emission, the dotted line represents ICs on the seed photons of synchrotron, BLR and MT, and the thick solid line represents total spectrum by summation all of emission components respectively. }
	\label{fig:3}
\end{figure*}

\begin{figure*}[ht!]
  \begin{center}
   \begin{tabular}{cc}
		\includegraphics[scale=0.2]{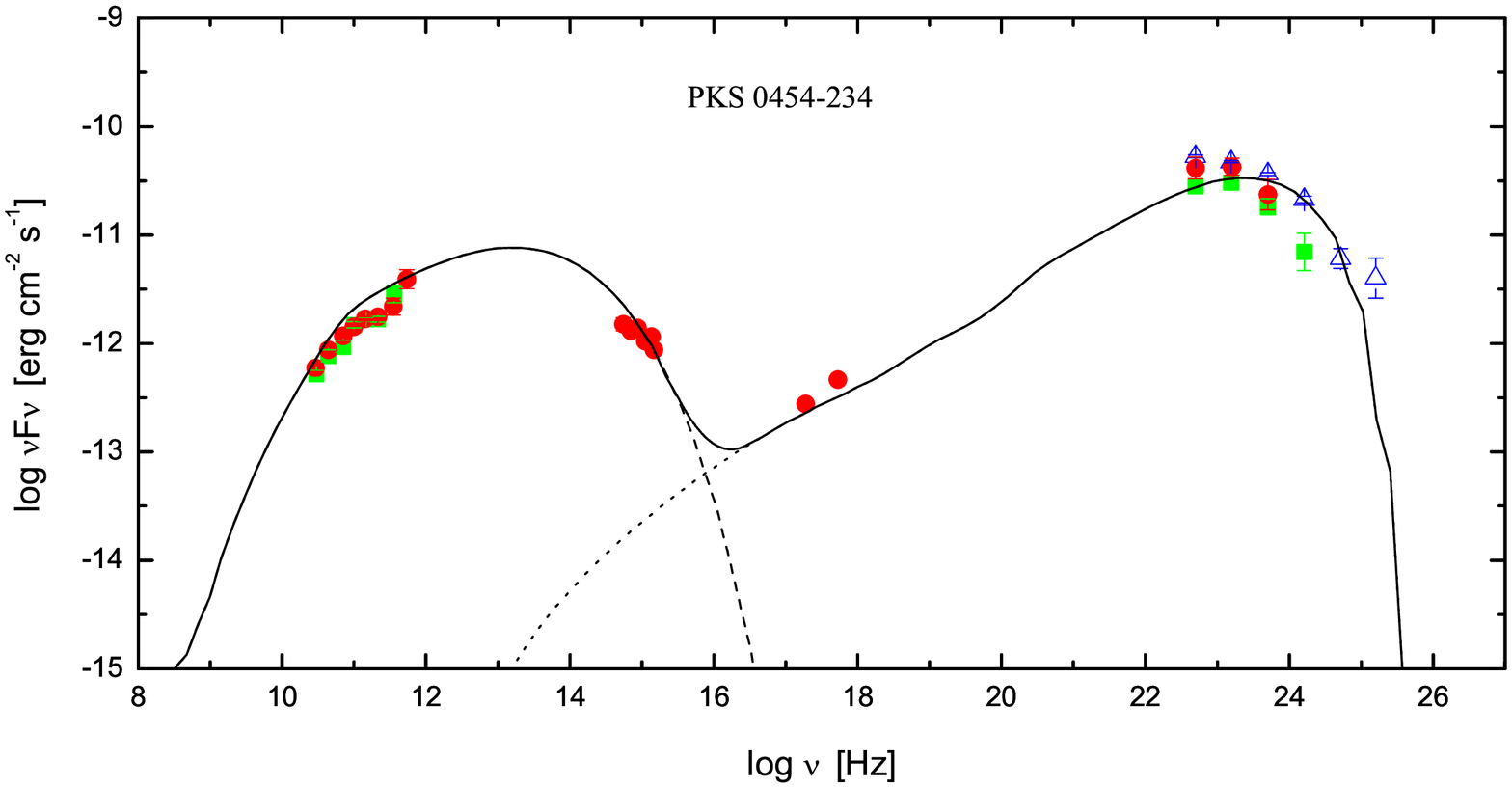}
        \includegraphics[scale=0.2]{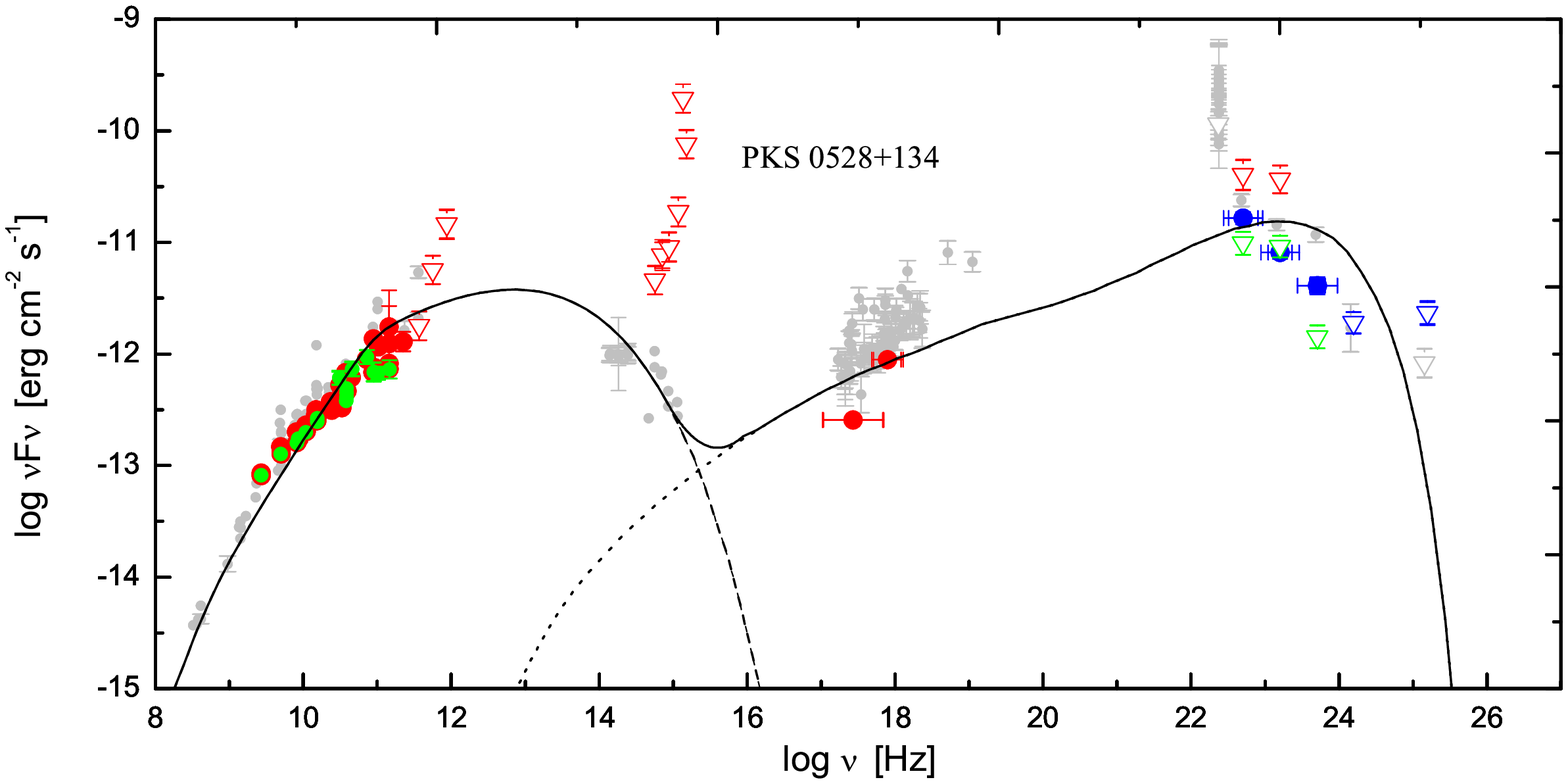}\\
        \includegraphics[scale=0.2]{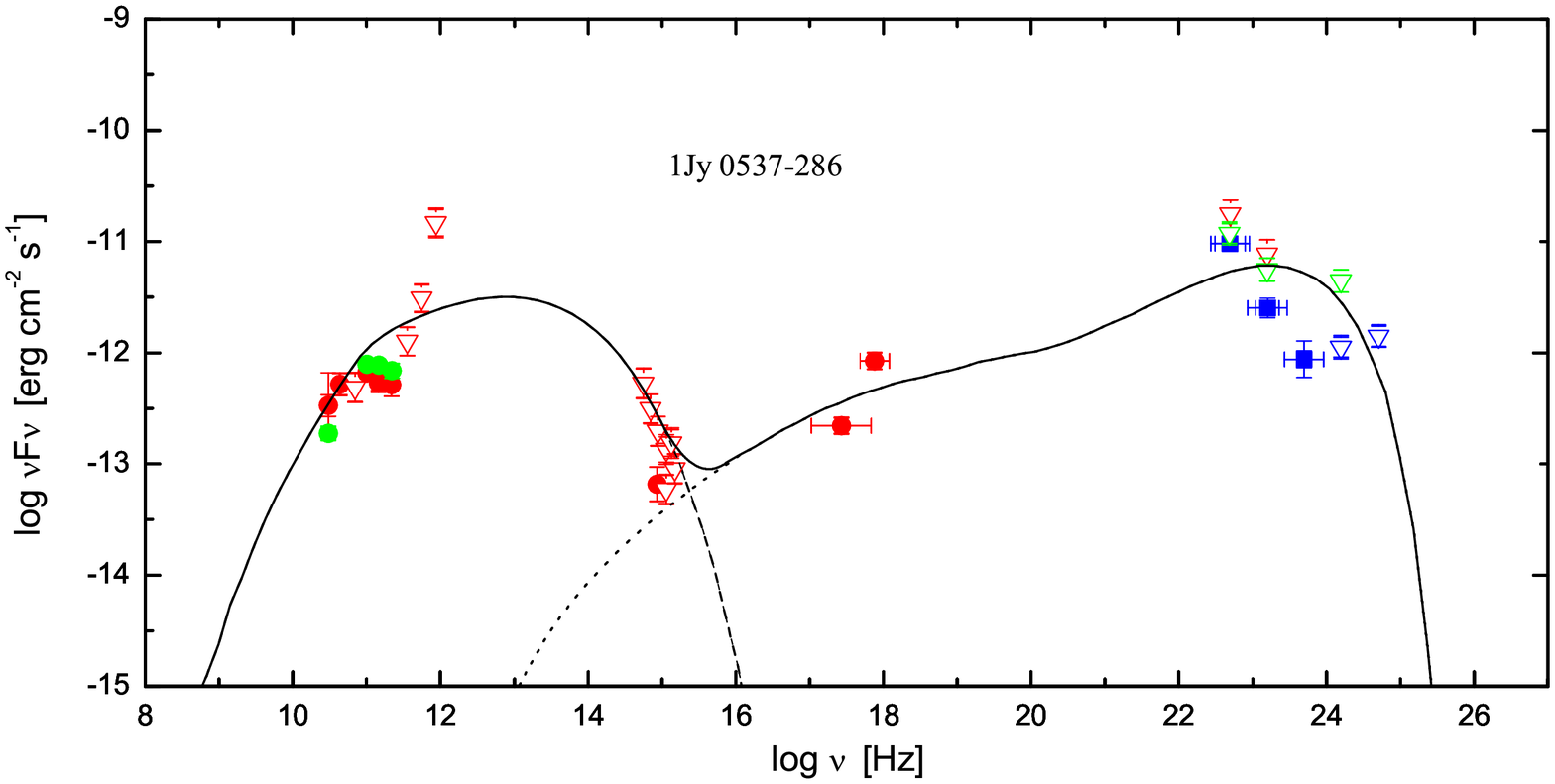}
        \includegraphics[scale=0.2]{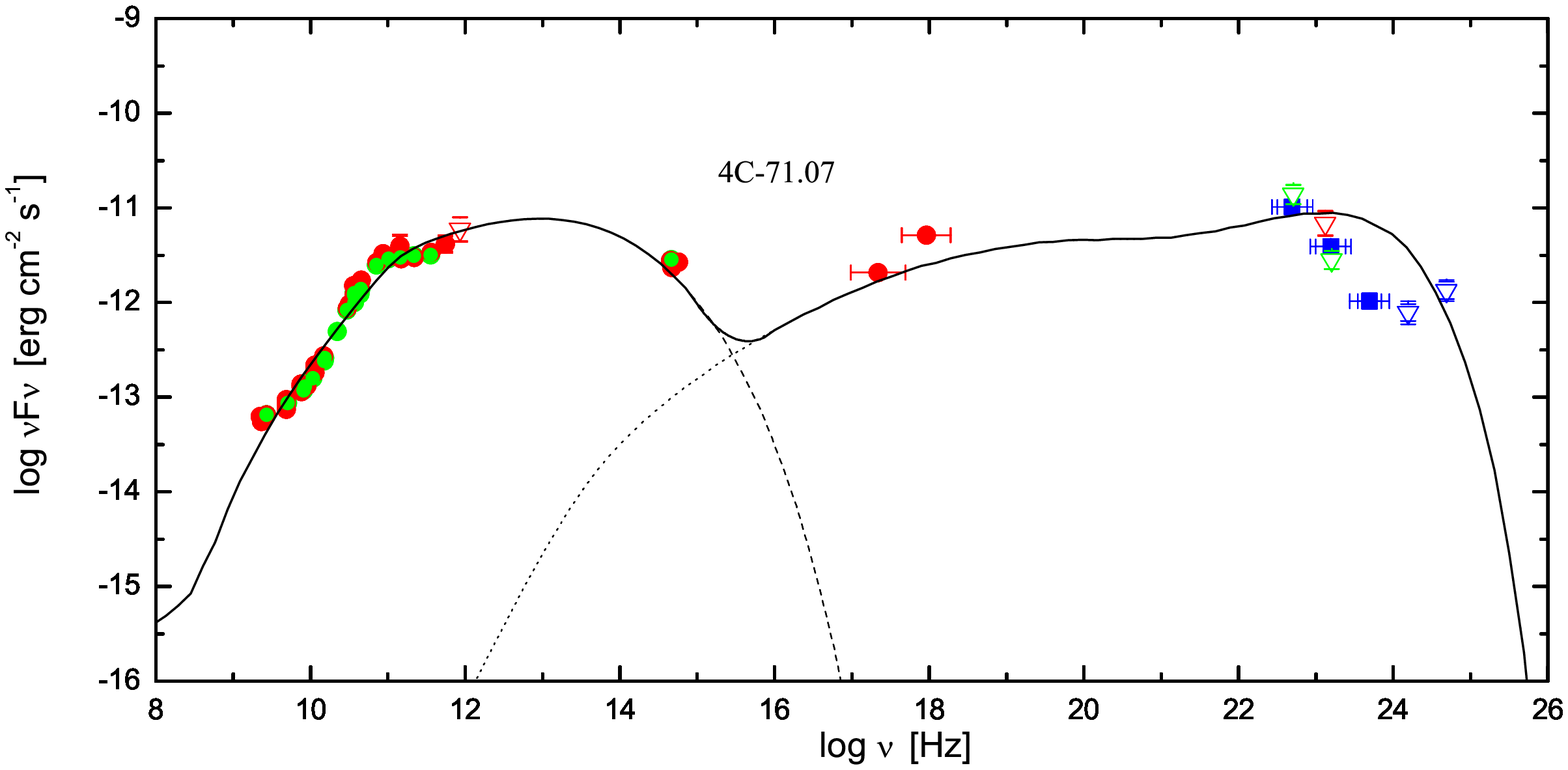}\\
        \includegraphics[scale=0.2]{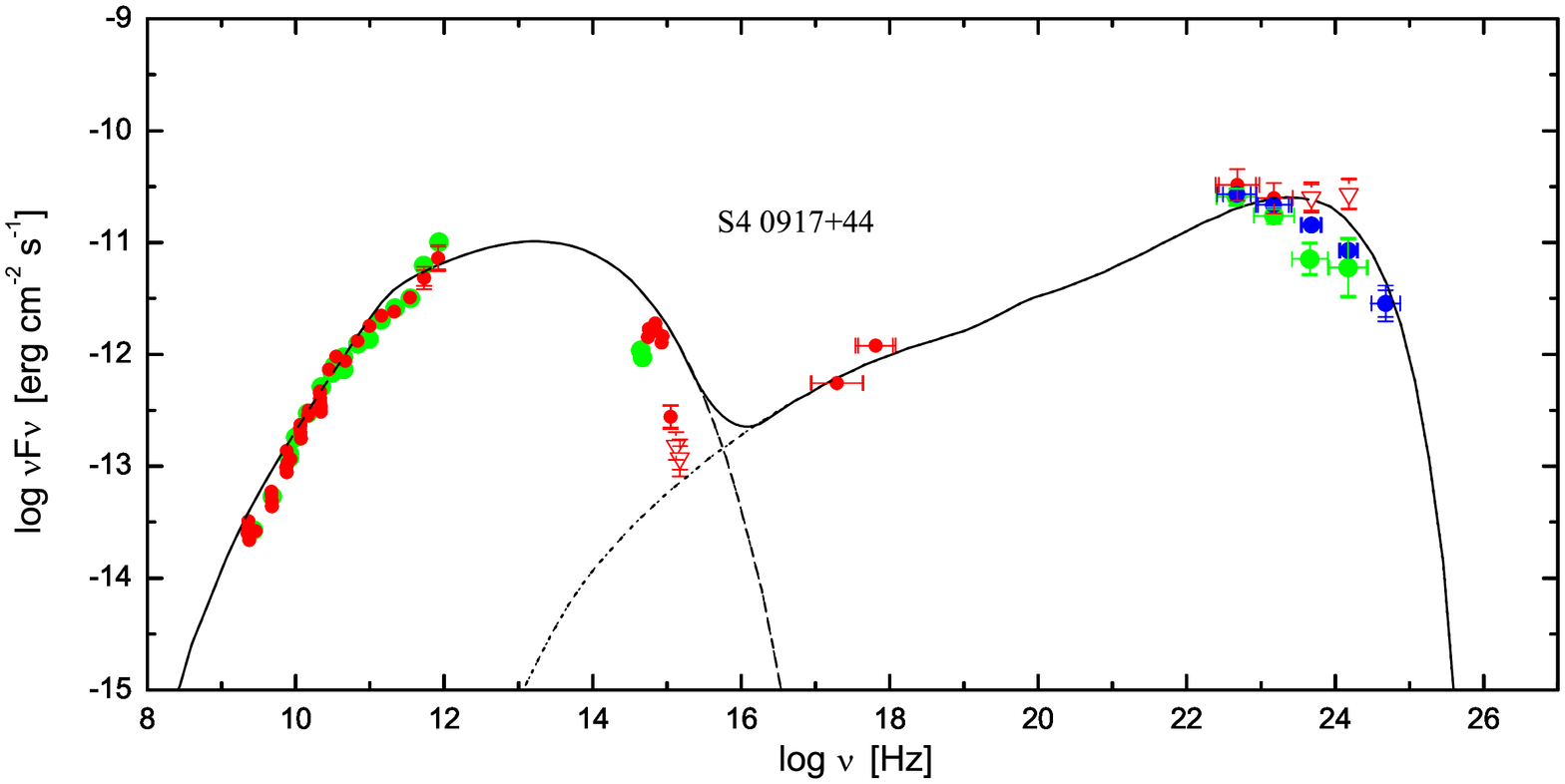}
        \includegraphics[scale=0.2]{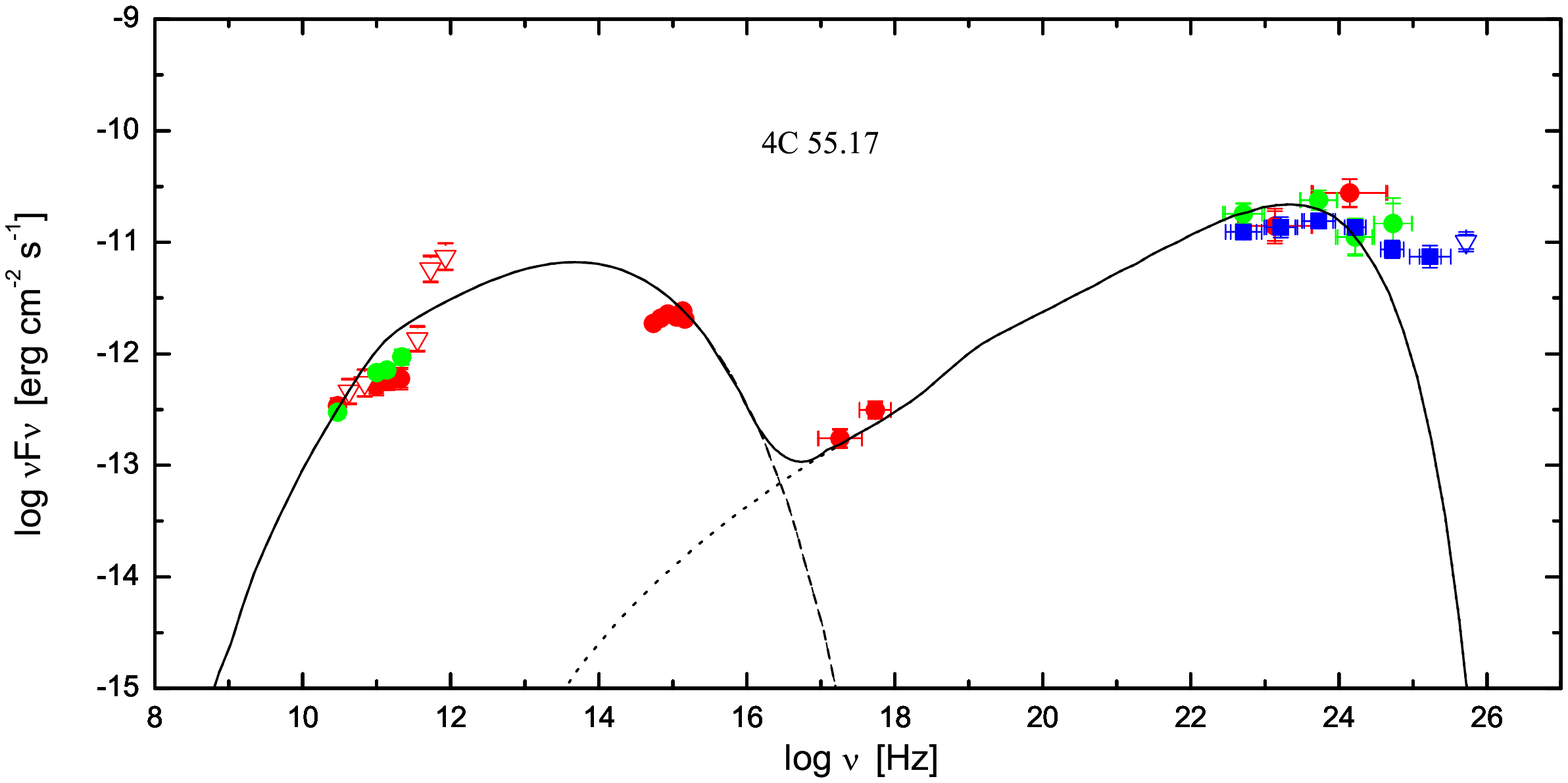}\\
   \end{tabular}
  \end{center}  	
	\caption{Comparisons of predicted multi-wavelength spectra with observed data for PKS 0454-234, PKS 0528-134, 1Jy 0537-286, 4C 71.07, S4 0917+44, 4C 55.17, respectively. Symbols and lines are same as in Fig. \ref{fig:3}.}
	\label{fig:4}
\end{figure*}

\begin{figure*}[ht!]
  \begin{center}
   \begin{tabular}{cc}
		\includegraphics[scale=0.2]{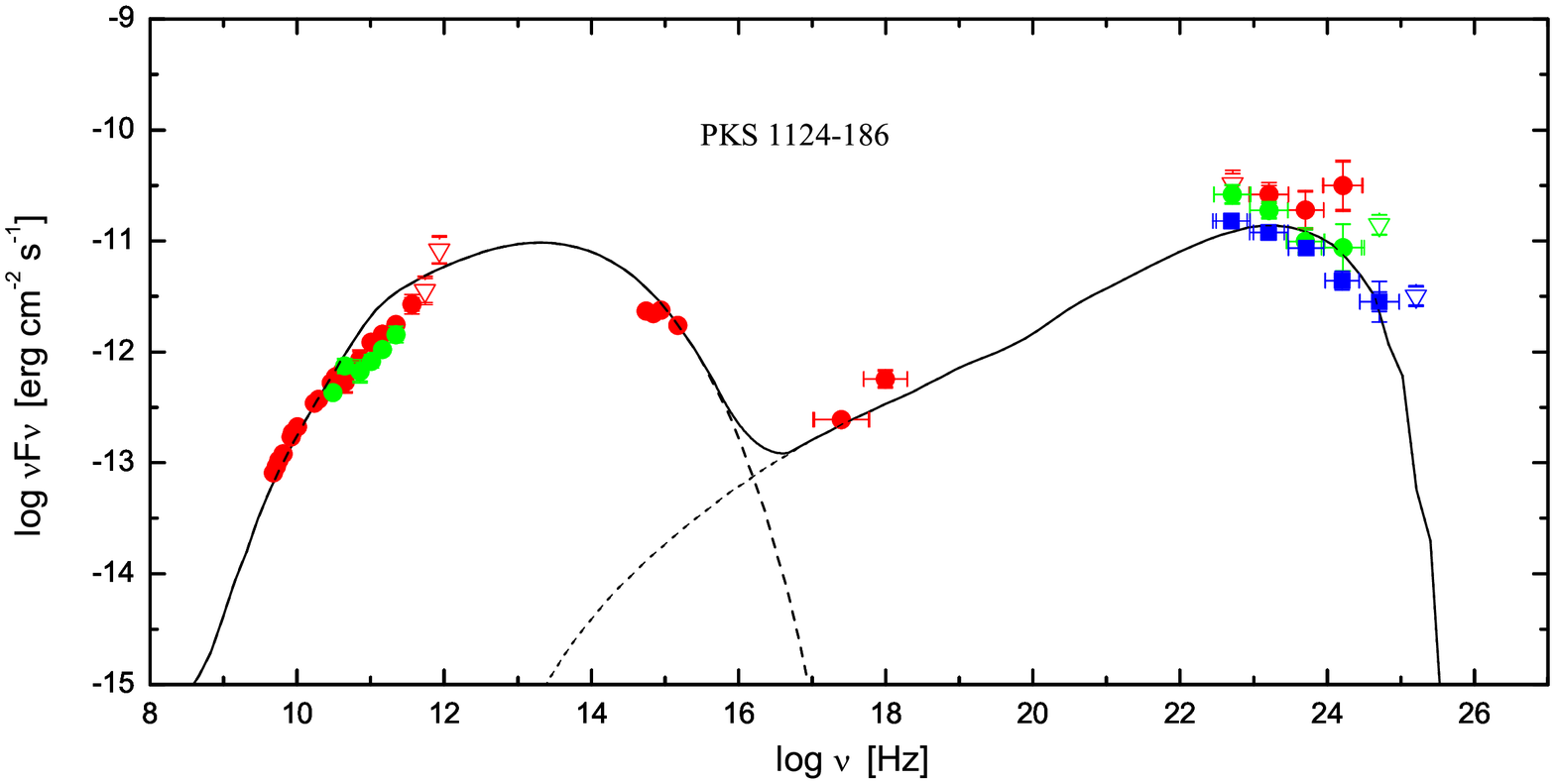}
        \includegraphics[scale=0.2]{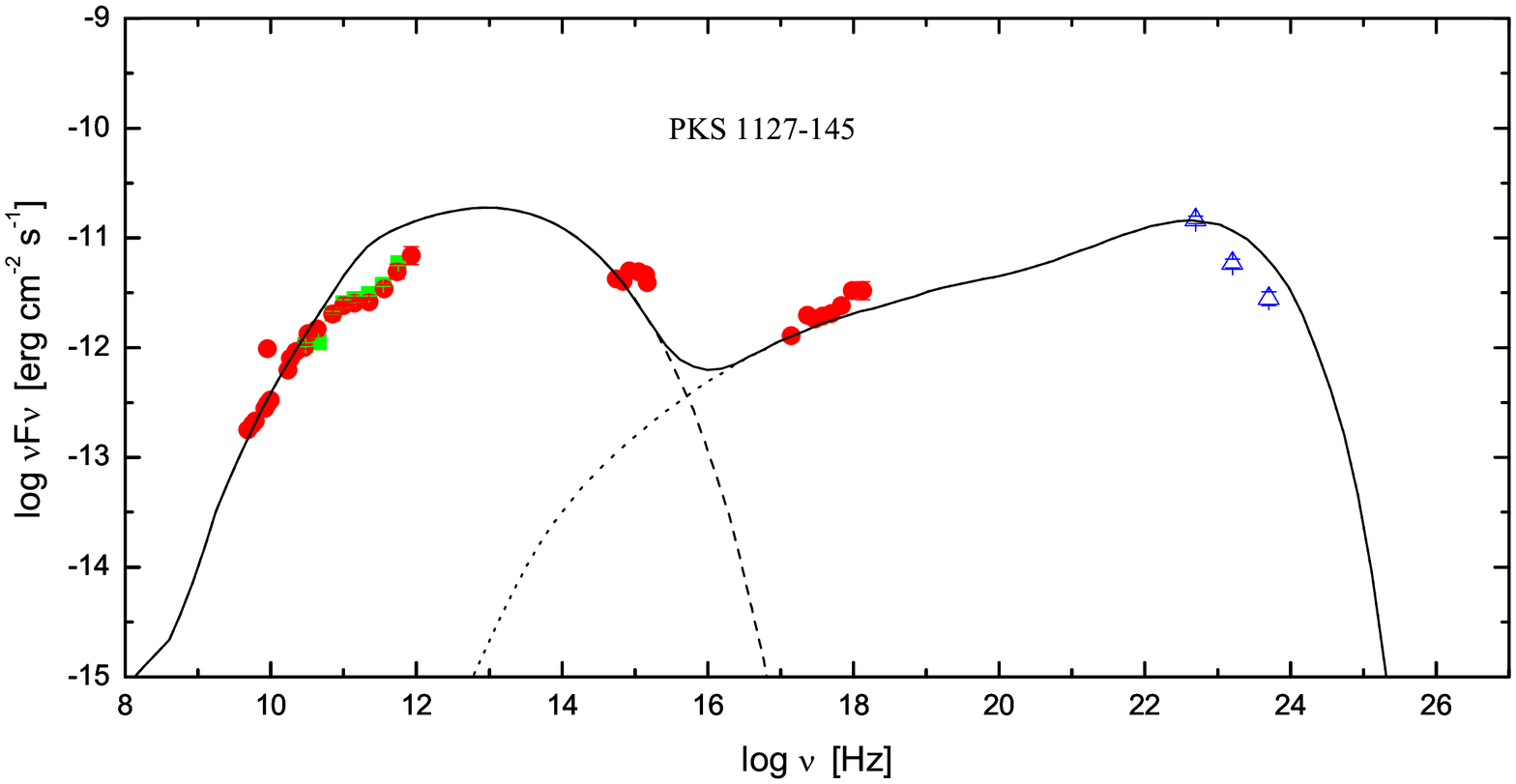}\\
        \includegraphics[scale=0.2]{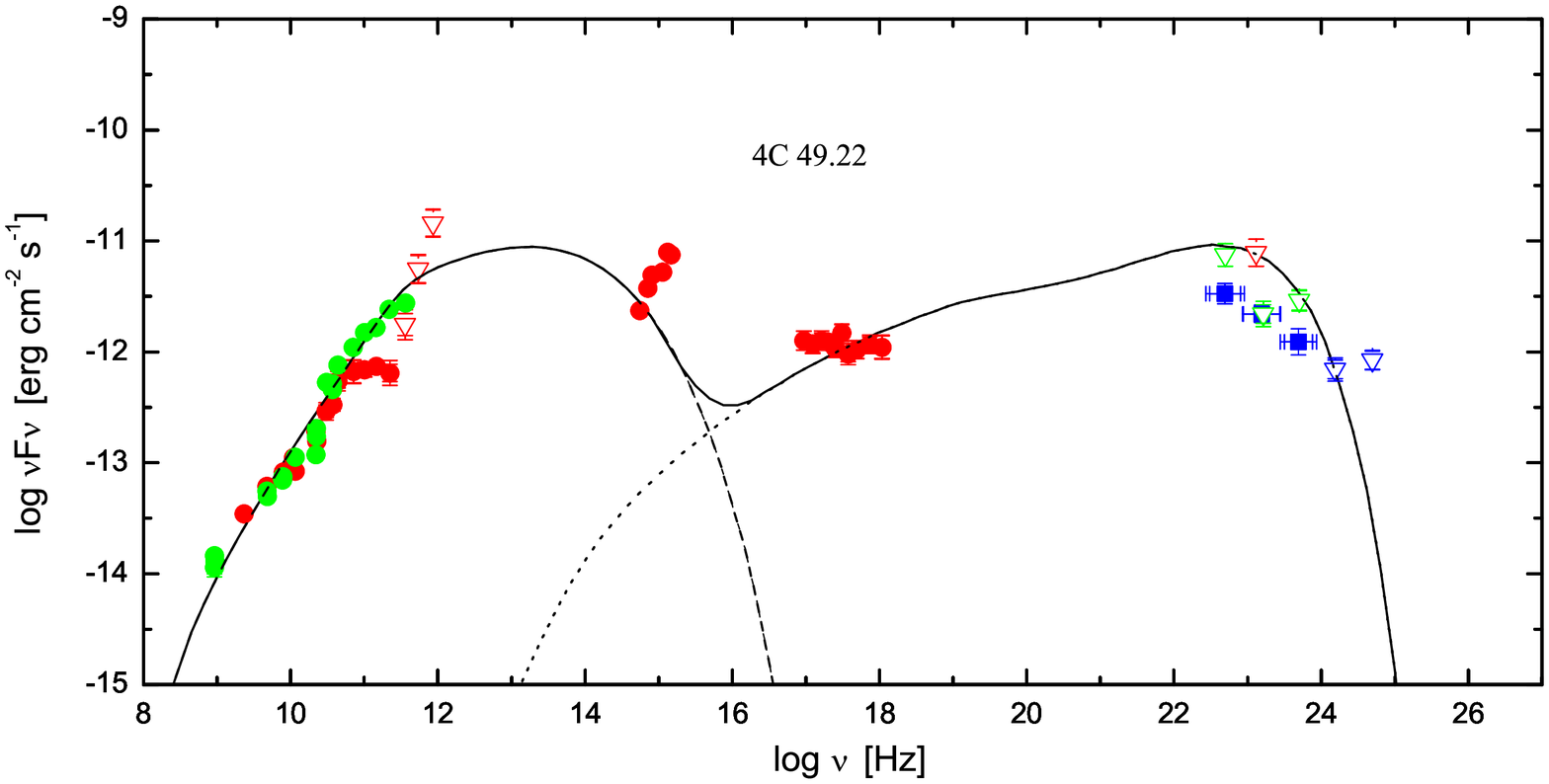}
        \includegraphics[scale=0.2]{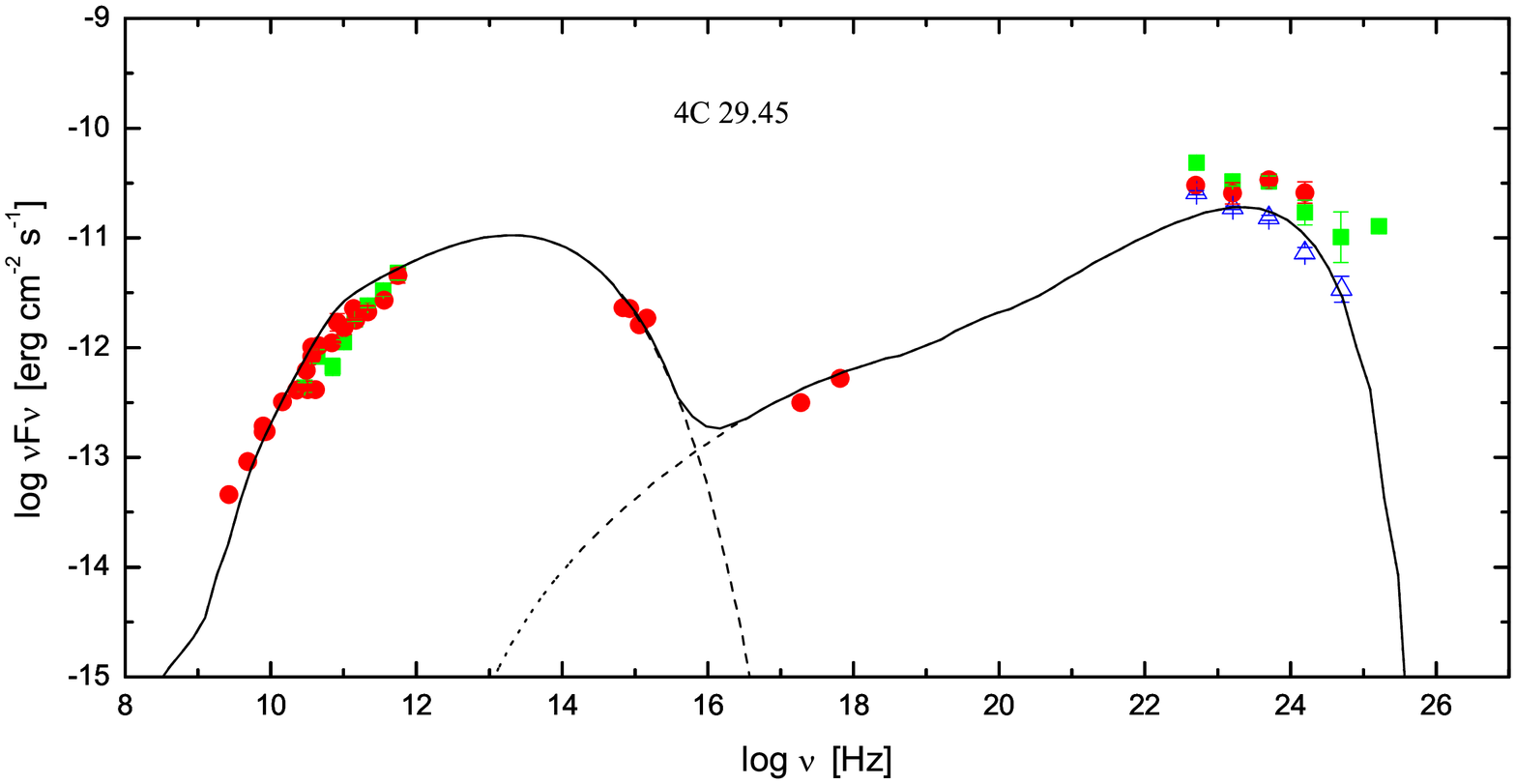}\\
        \includegraphics[scale=0.2]{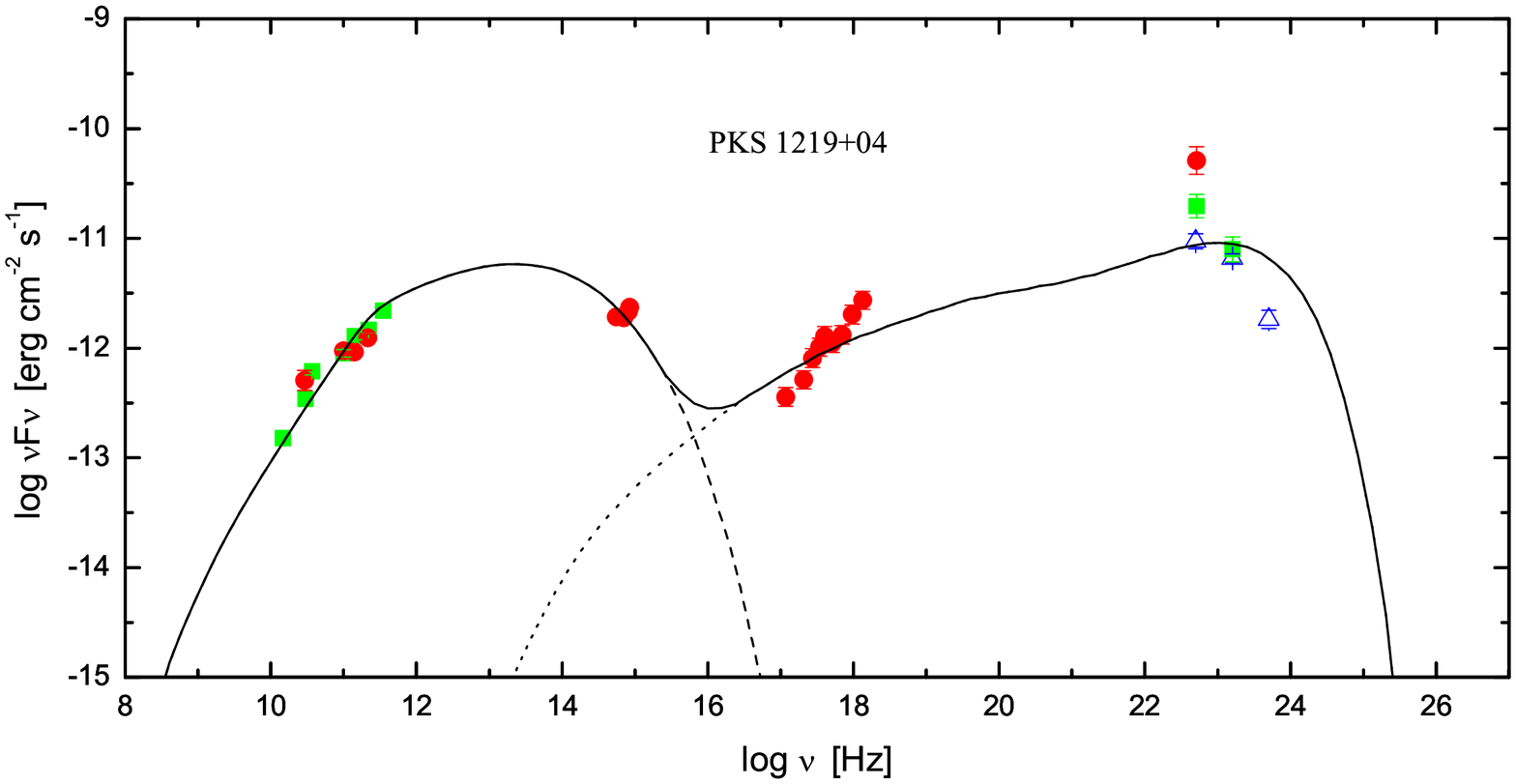}
        \includegraphics[scale=0.2]{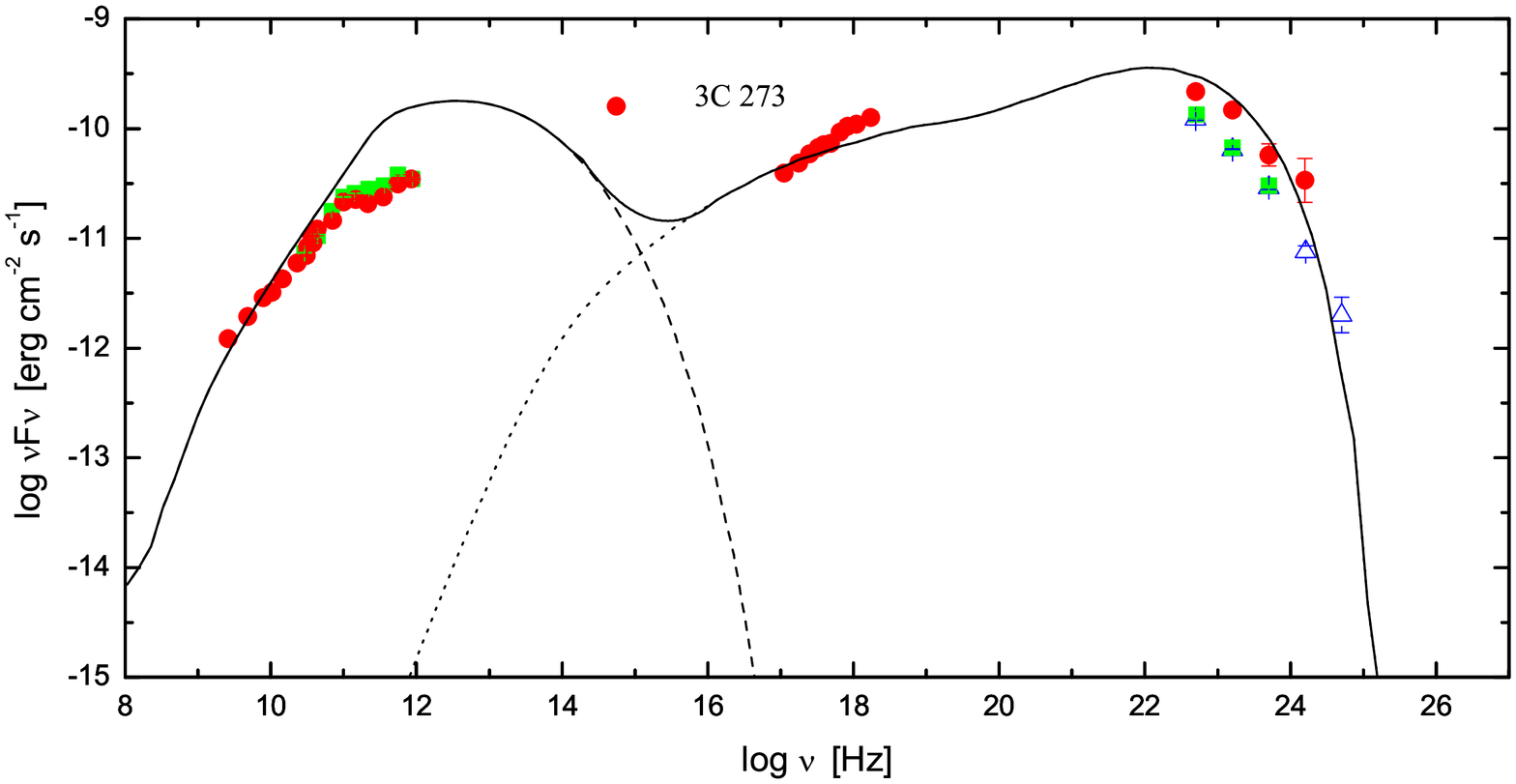}\\
    \end{tabular}
  \end{center}     		
	\caption{Comparisons of predicted multi-wavelength spectra with observed data for PKS 1124-186, PKS 1127-145, 4C 49.22, 4C 29.45, PKS 1219+04, 3C 273, respectively. Symbols and lines are same as in Fig. \ref{fig:3}.}
	\label{fig:5}
\end{figure*}

\begin{figure*}[ht!]
  \begin{center}
   \begin{tabular}{cc}
        \includegraphics[scale=0.2]{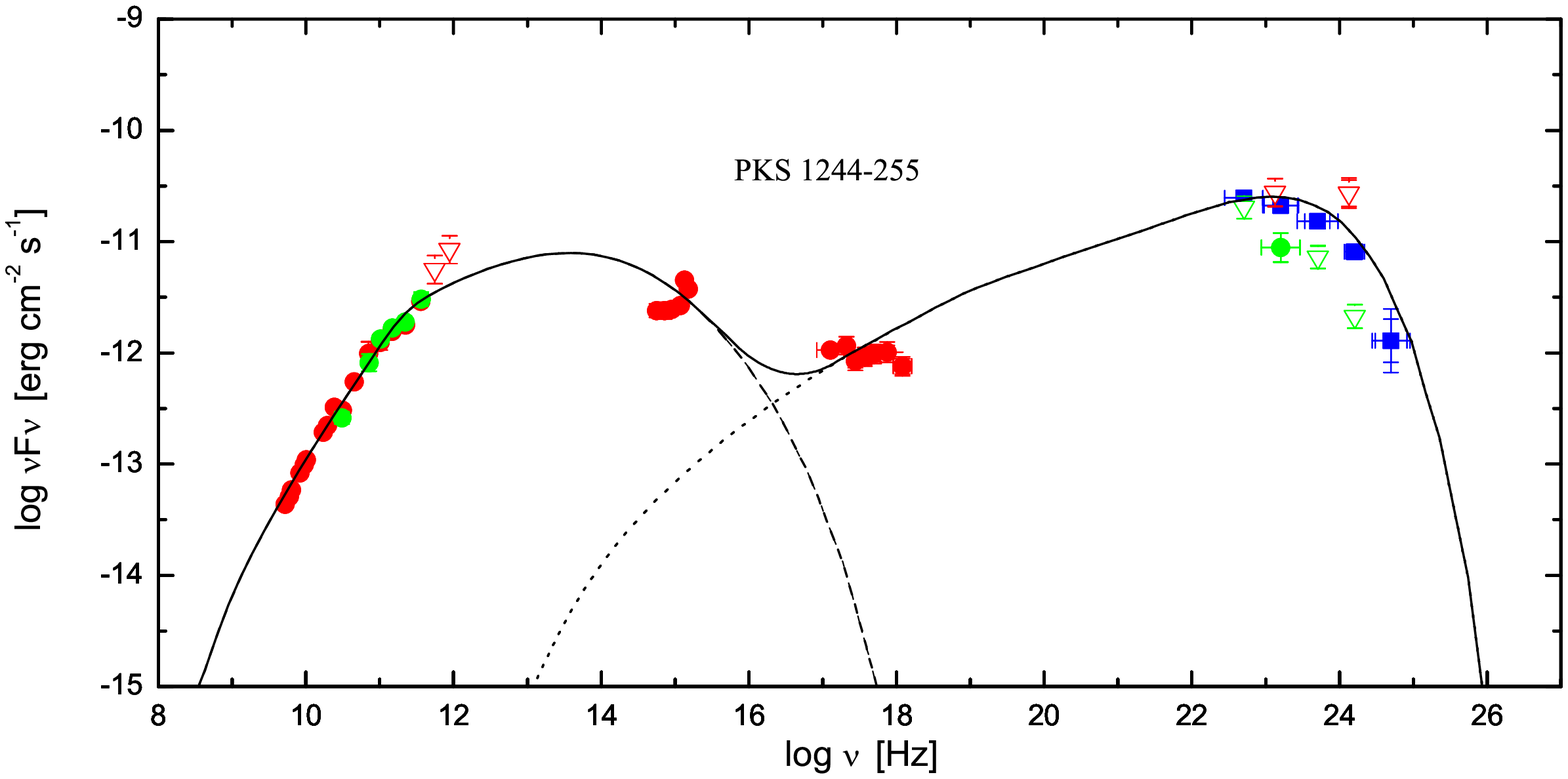}
        \includegraphics[scale=0.2]{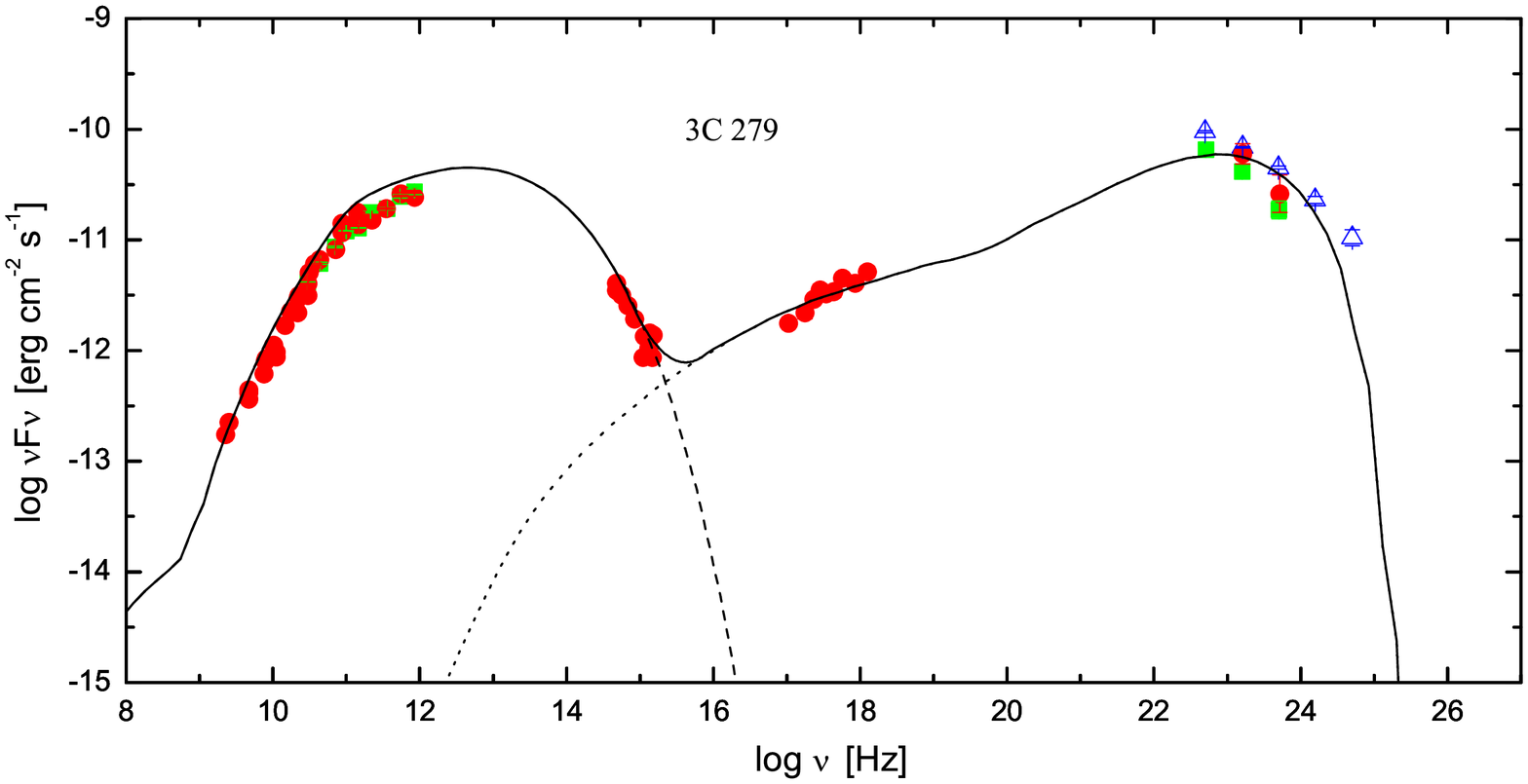}\\
        \includegraphics[scale=0.2]{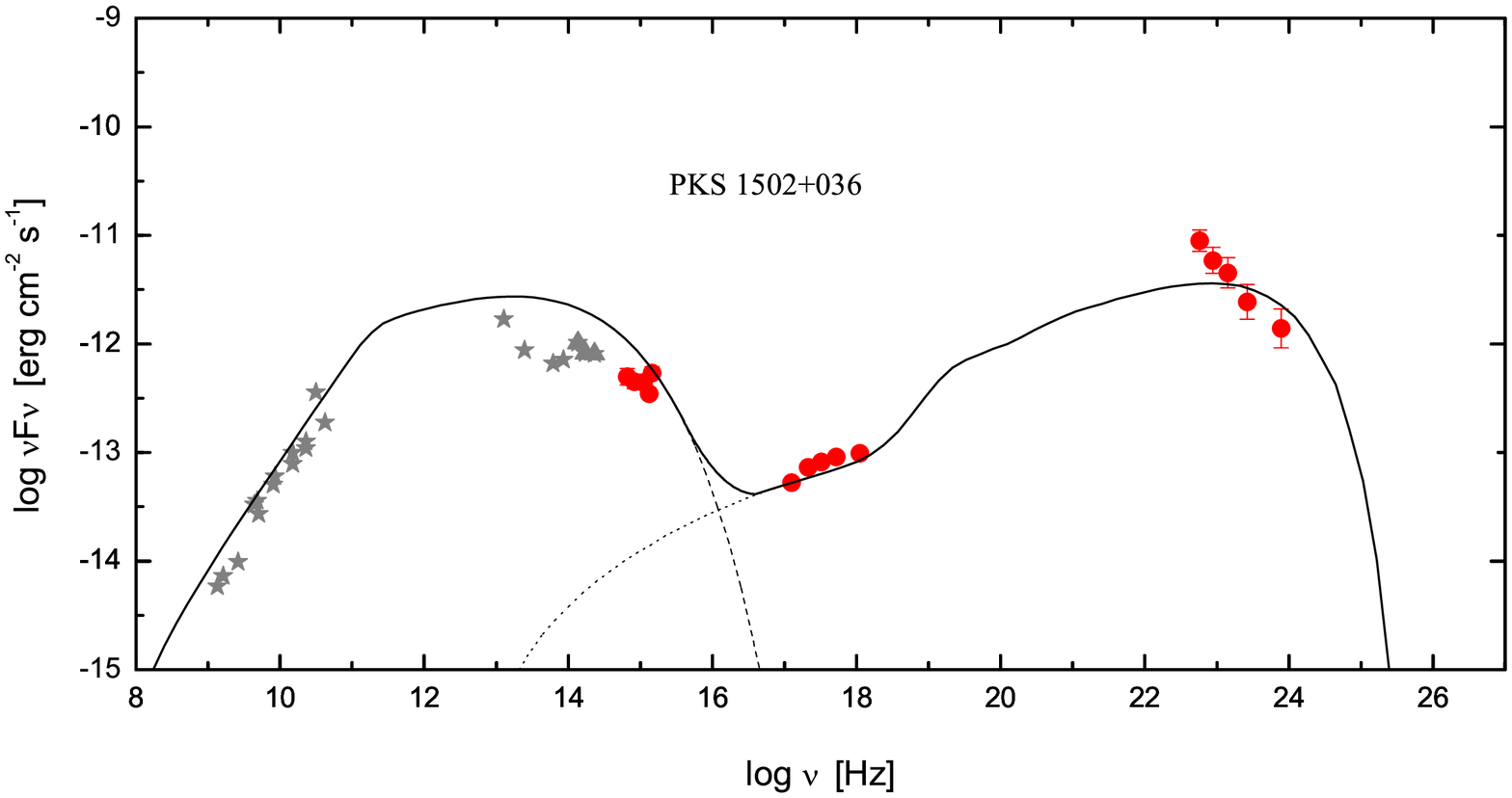}
        \includegraphics[scale=0.2]{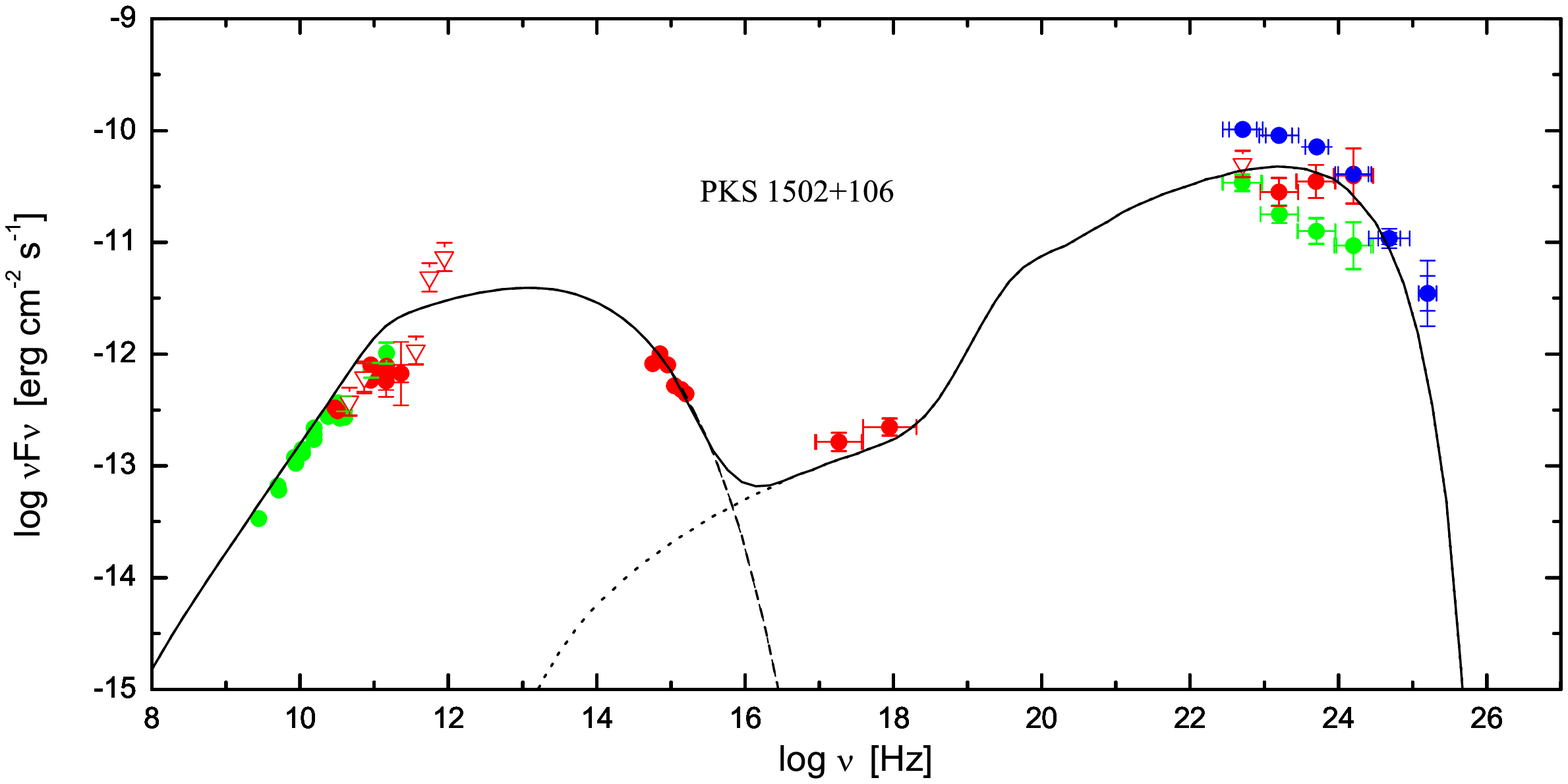}\\
        \includegraphics[scale=0.2]{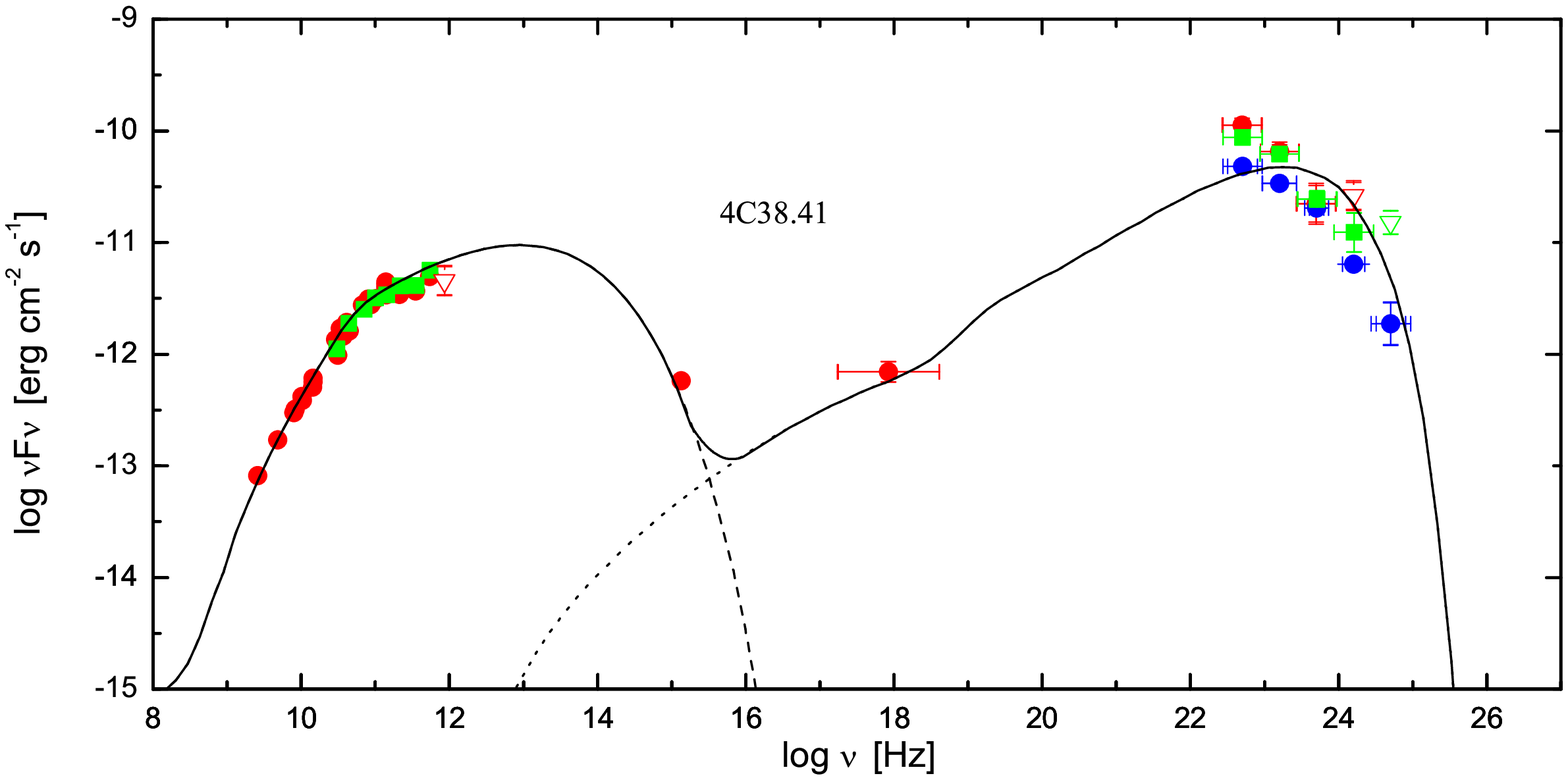}
        \includegraphics[scale=0.2]{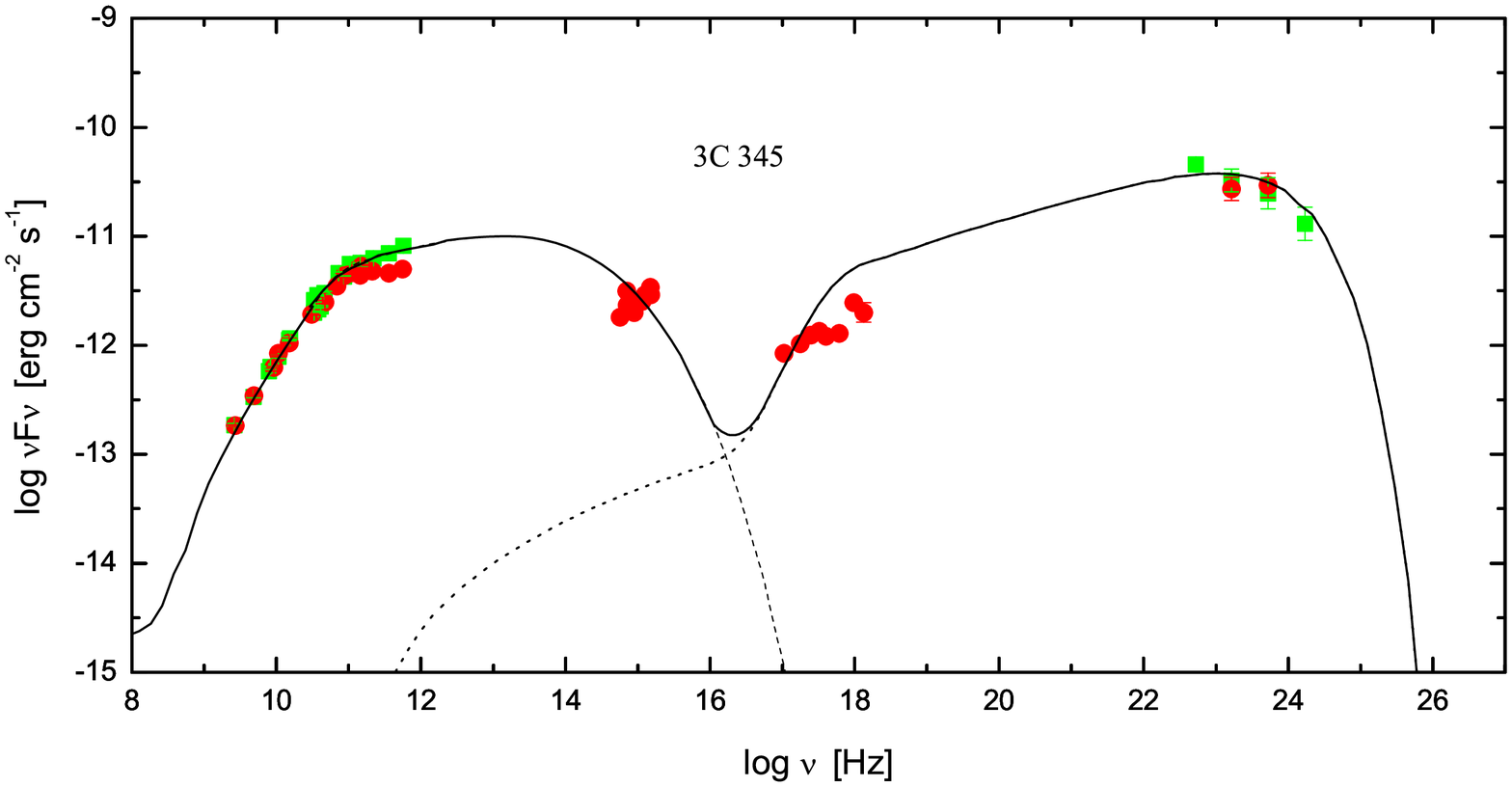}\\
    \end{tabular}
  \end{center}       		
	\caption{Comparisons of predicted multi-wavelength spectra with observed data for PKS 1244-255, 3C 279, PKS 1502+036, PKS 1502+106, 4C 38.41, 3C 345, respectively. Symbols and lines are same as in Fig. \ref{fig:3}.}
	\label{fig:6}
\end{figure*}

\begin{figure*}[ht!]
  \begin{center}
   \begin{tabular}{cc}
		\includegraphics[scale=0.2]{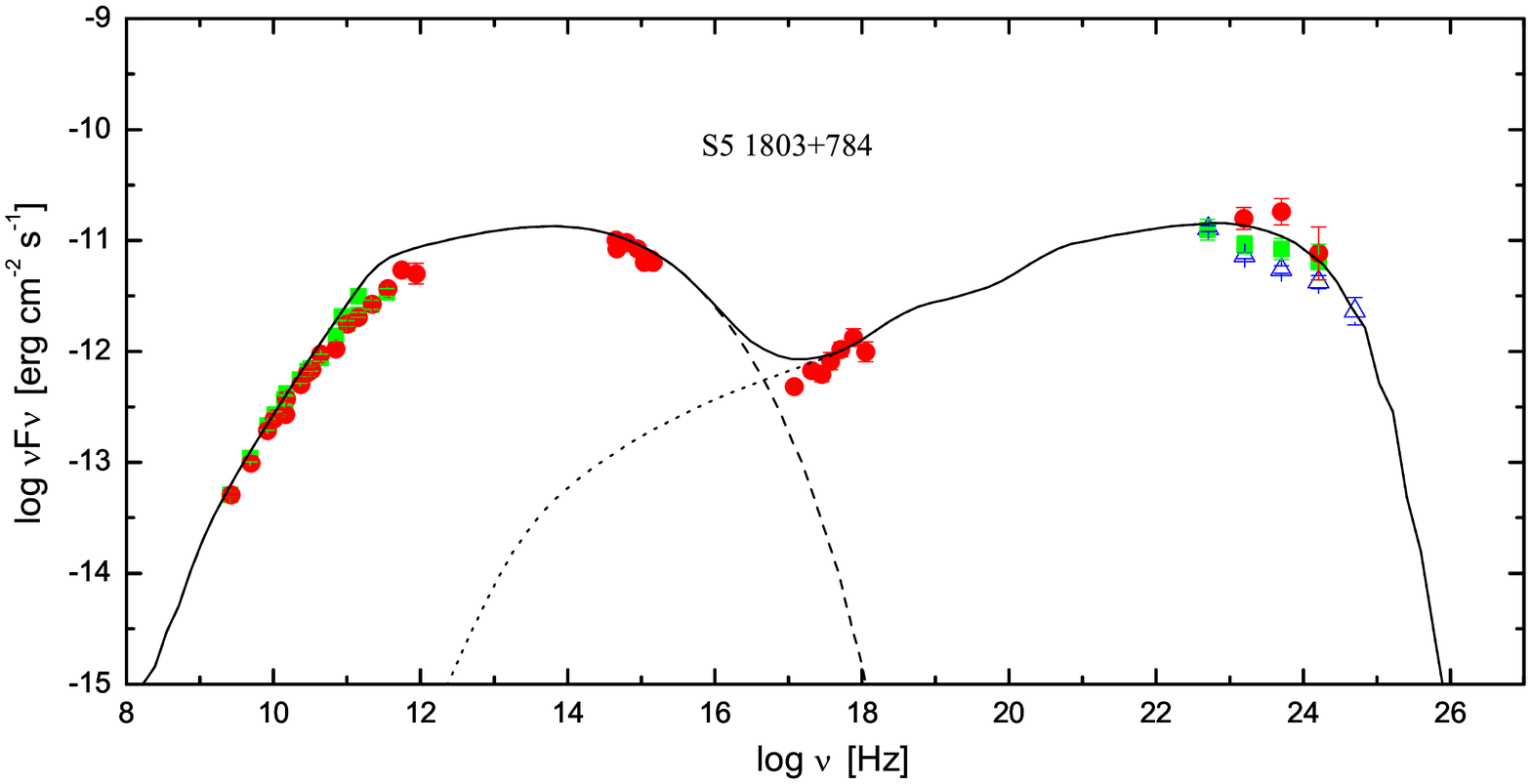}
        \includegraphics[scale=0.2]{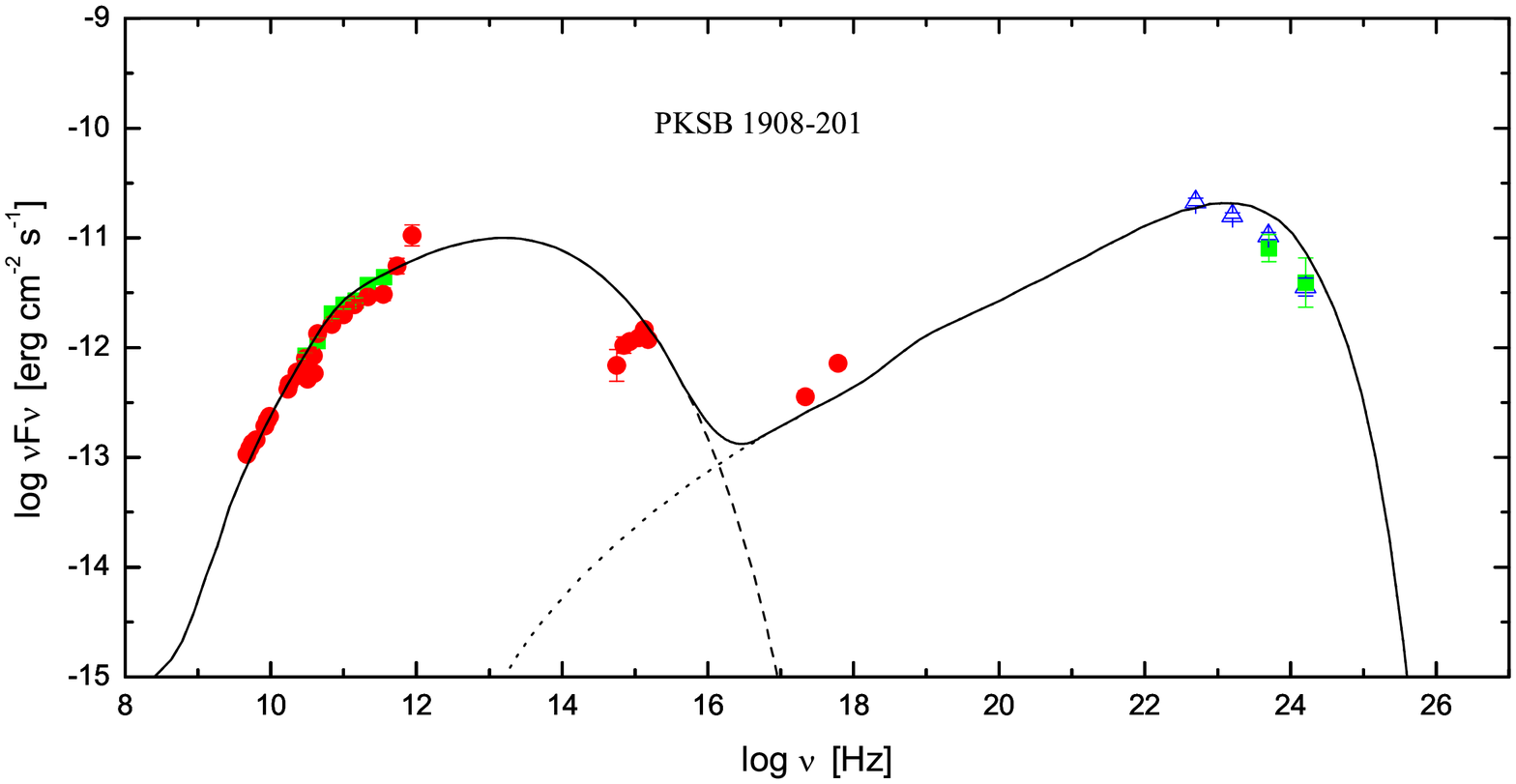}\\
        \includegraphics[scale=0.2]{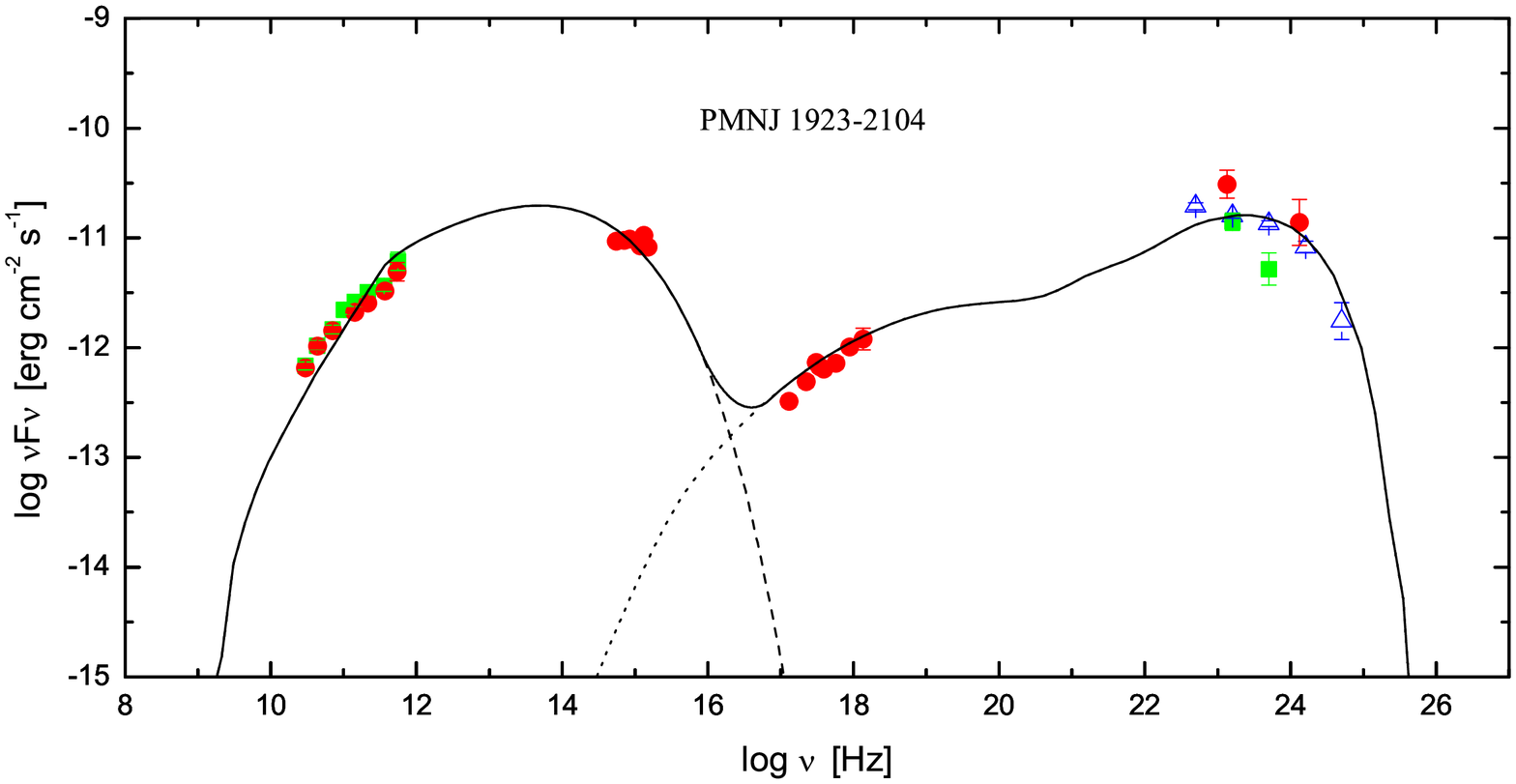}
        \includegraphics[scale=0.2]{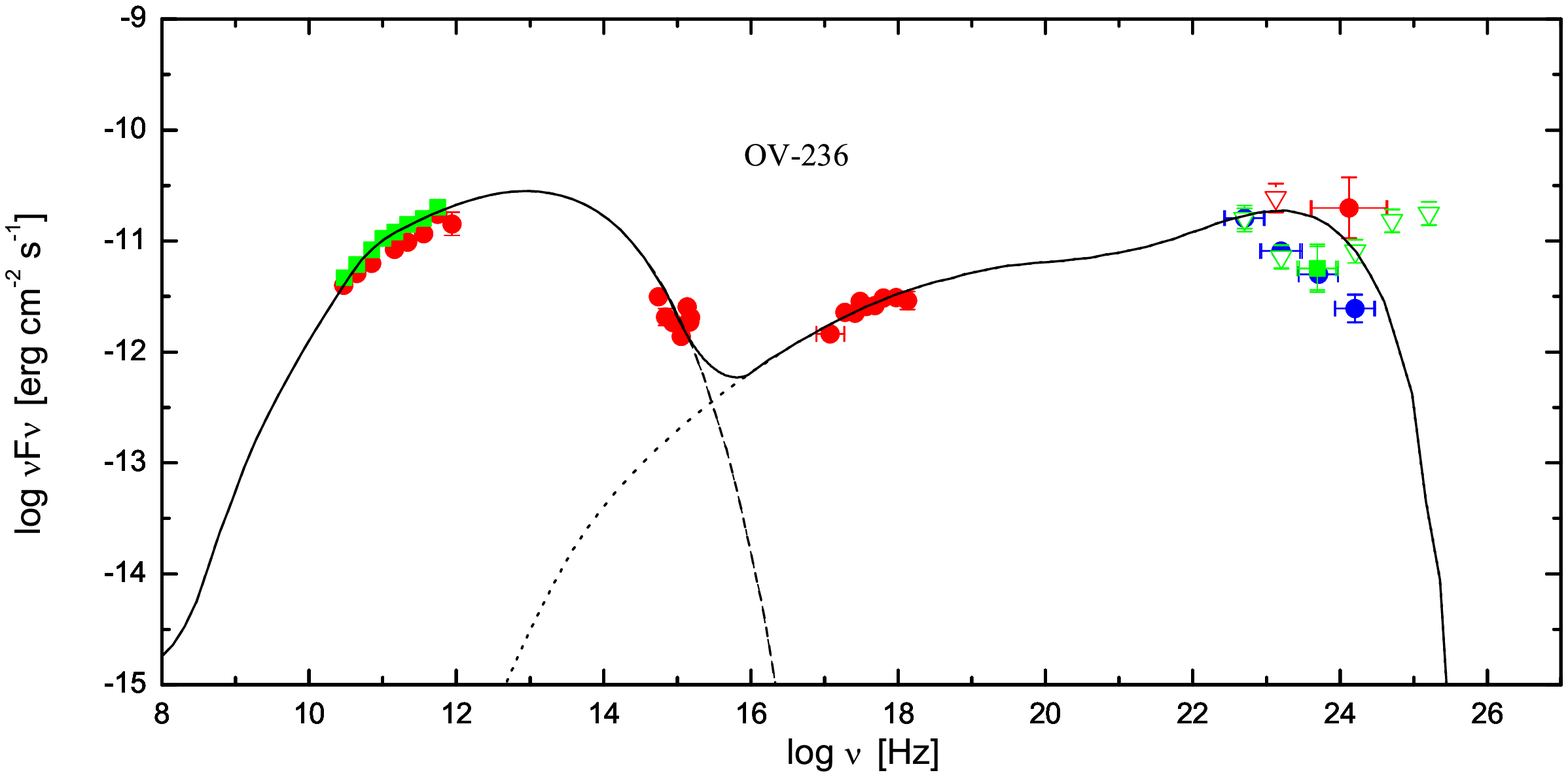}\\
        \includegraphics[scale=0.2]{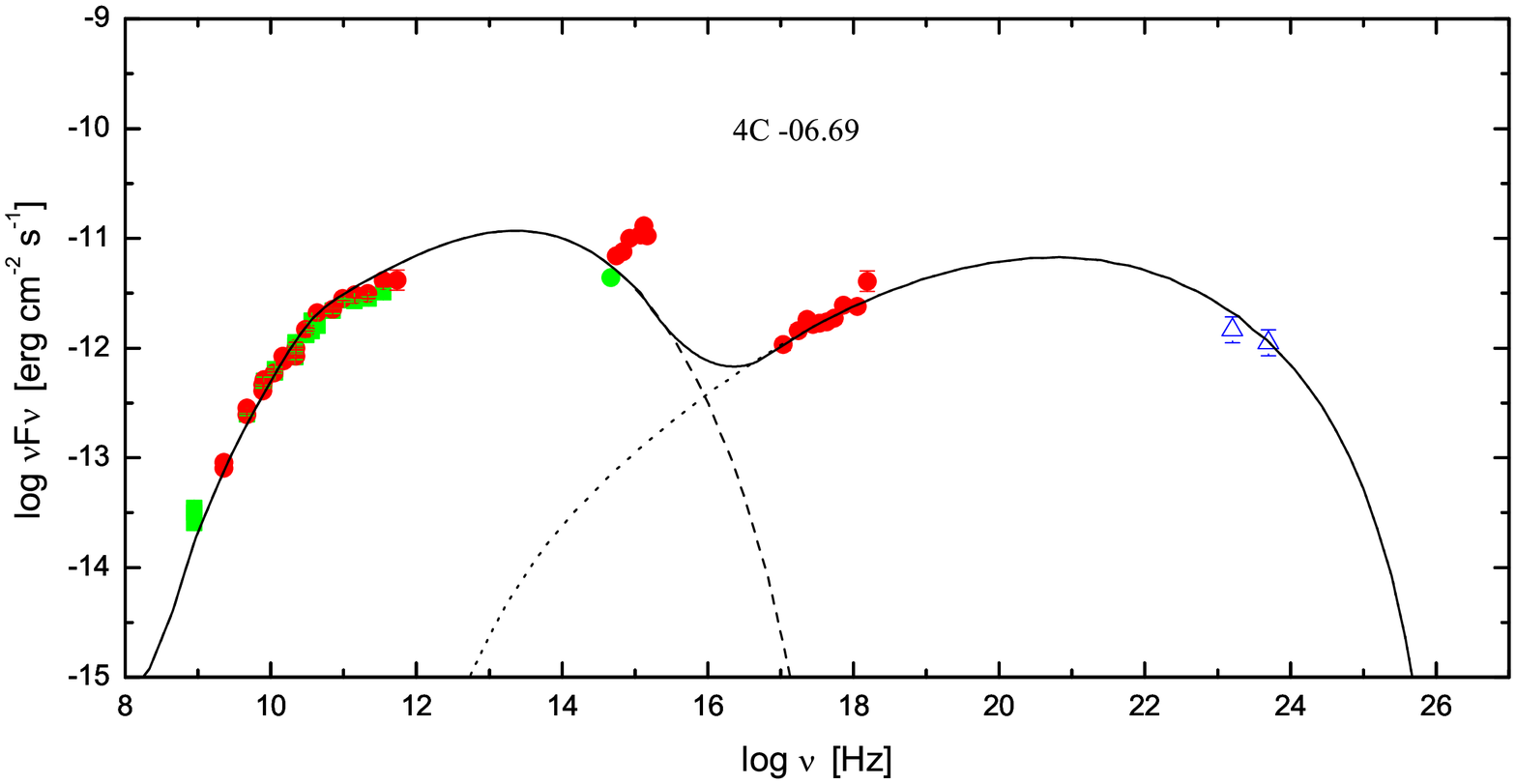}
        \includegraphics[scale=0.2]{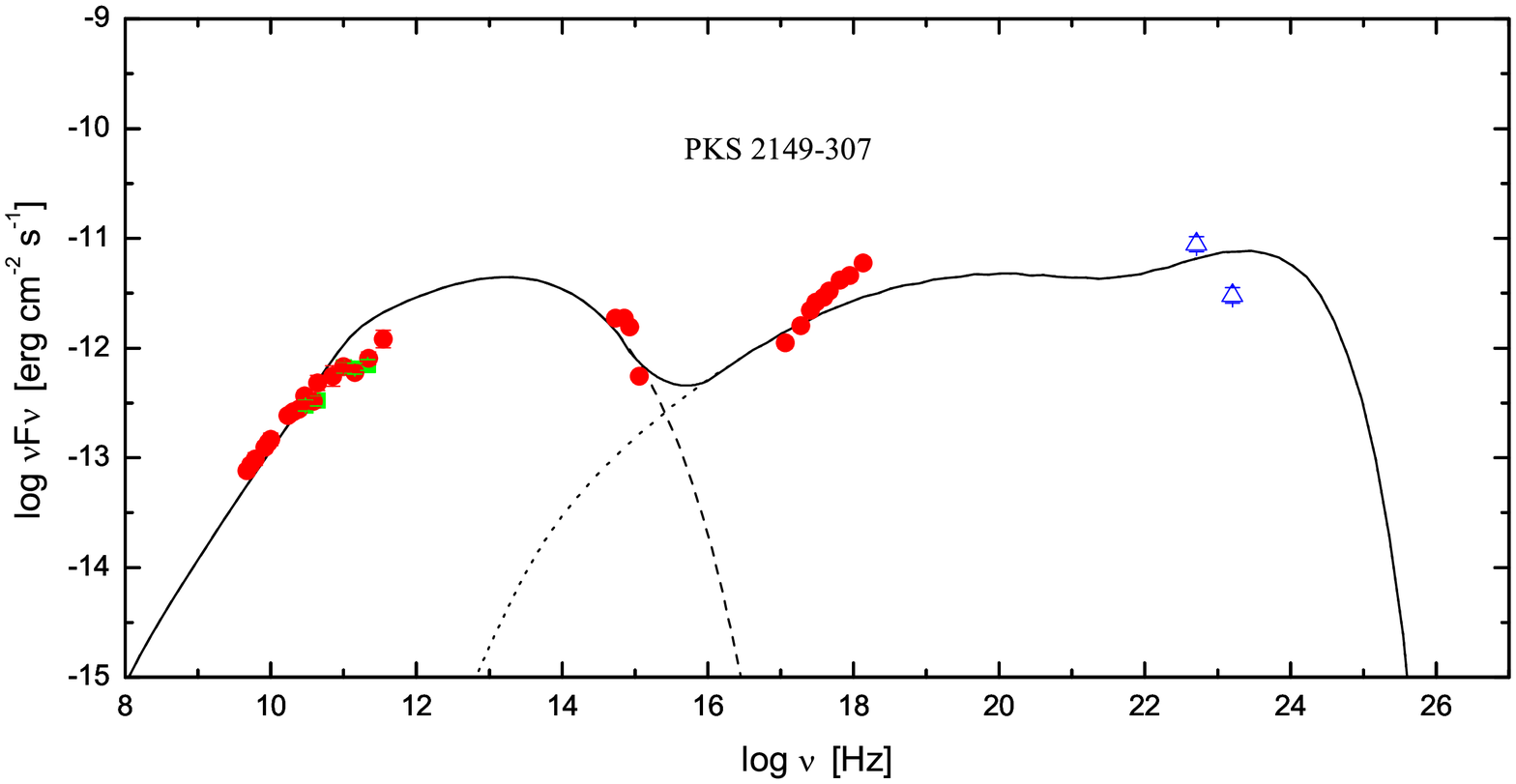}\\
     \end{tabular}
  \end{center}     		
	\caption{Comparisons of predicted multi-wavelength spectra with observed data for S5 1803+784, PKSB 1908-201, PMNJ 1923-2104, OV-236, 4C -06.69, PKS 2149-307, respectively. Symbols and lines are same as in Fig. \ref{fig:3}.}
	\label{fig:7}
\end{figure*}

\begin{figure*}[ht!]
  \begin{center}
   \begin{tabular}{cc}
		
        \includegraphics[scale=0.2]{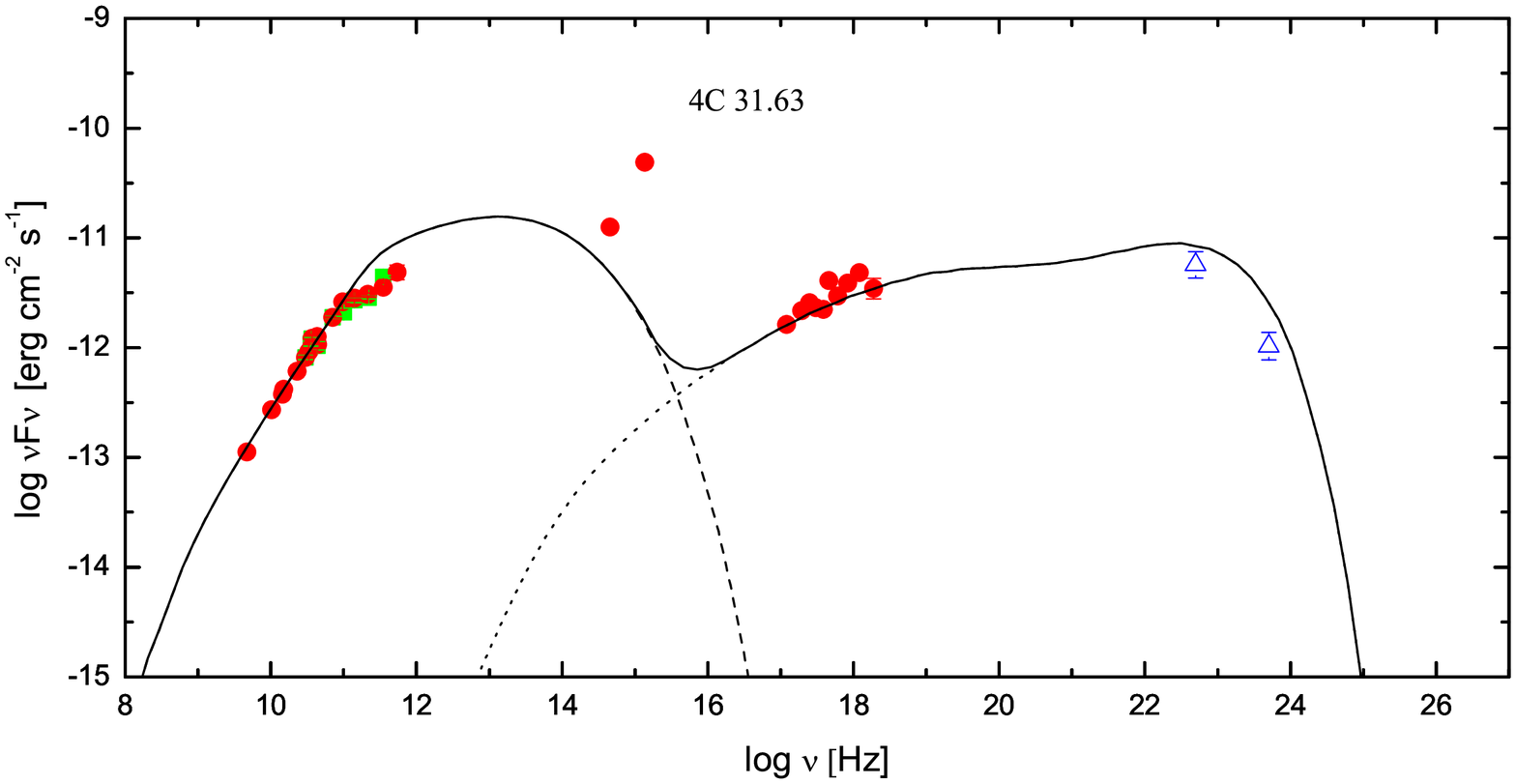}
        \includegraphics[scale=0.2]{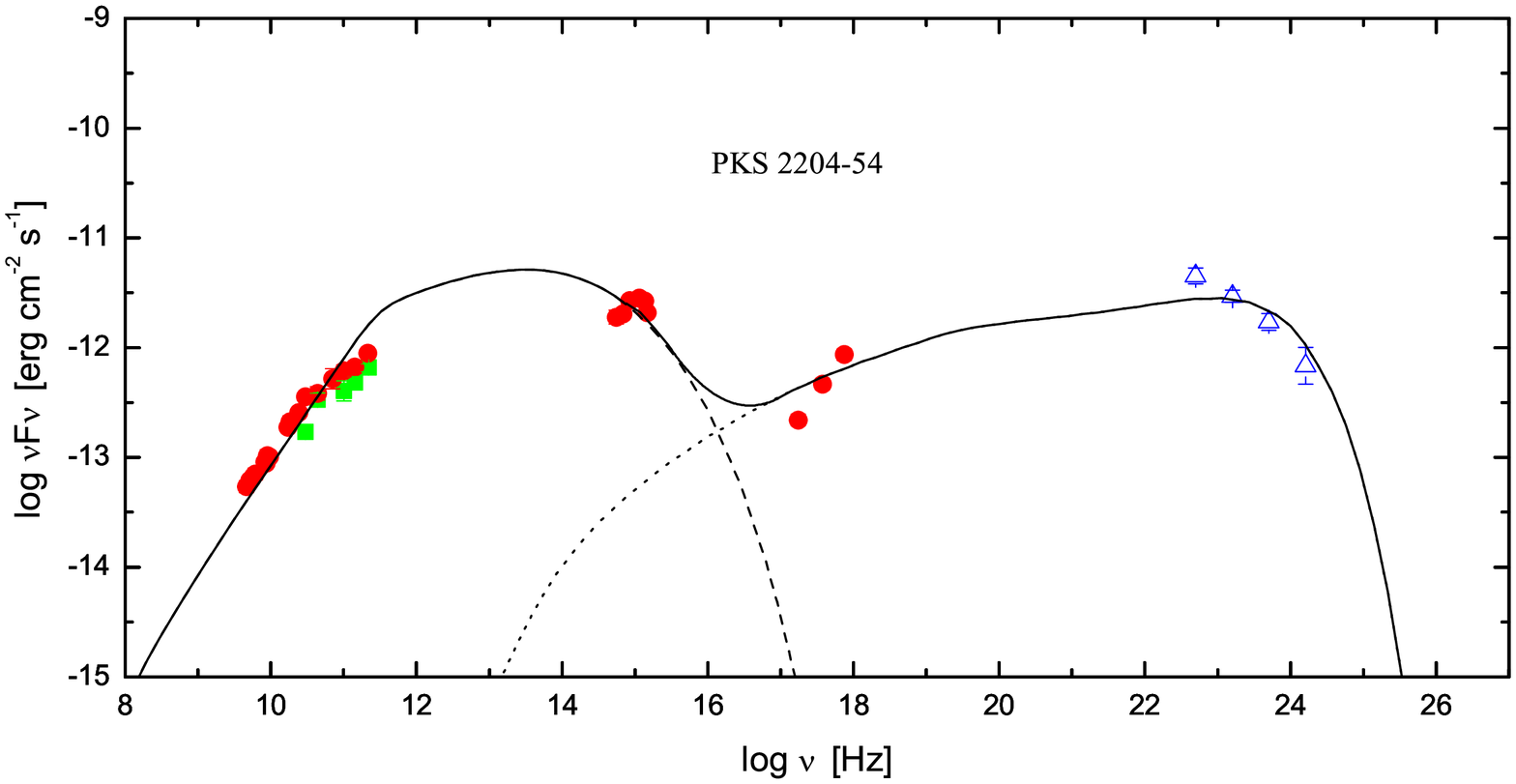}\\
        \includegraphics[scale=0.2]{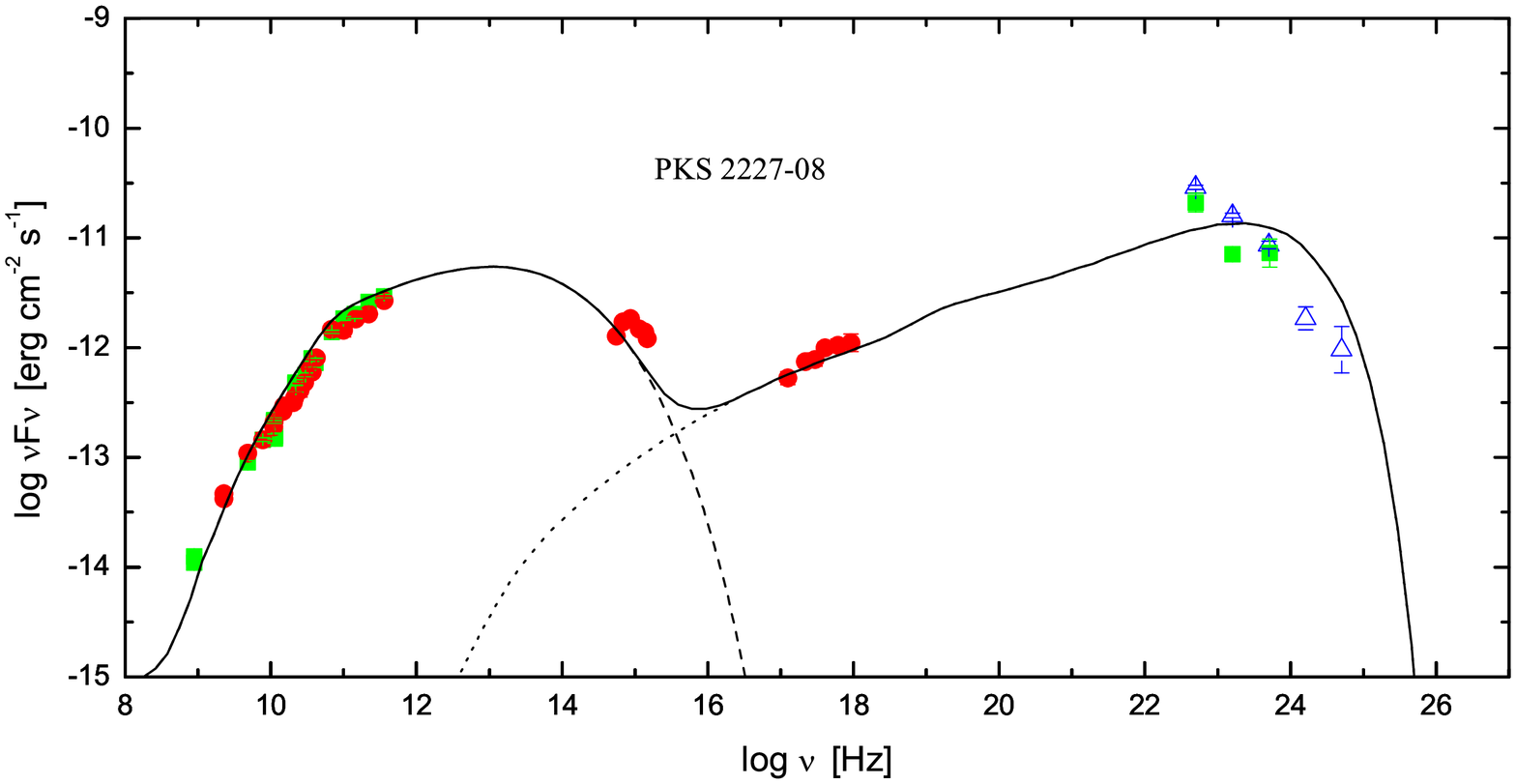}
        \includegraphics[scale=0.2]{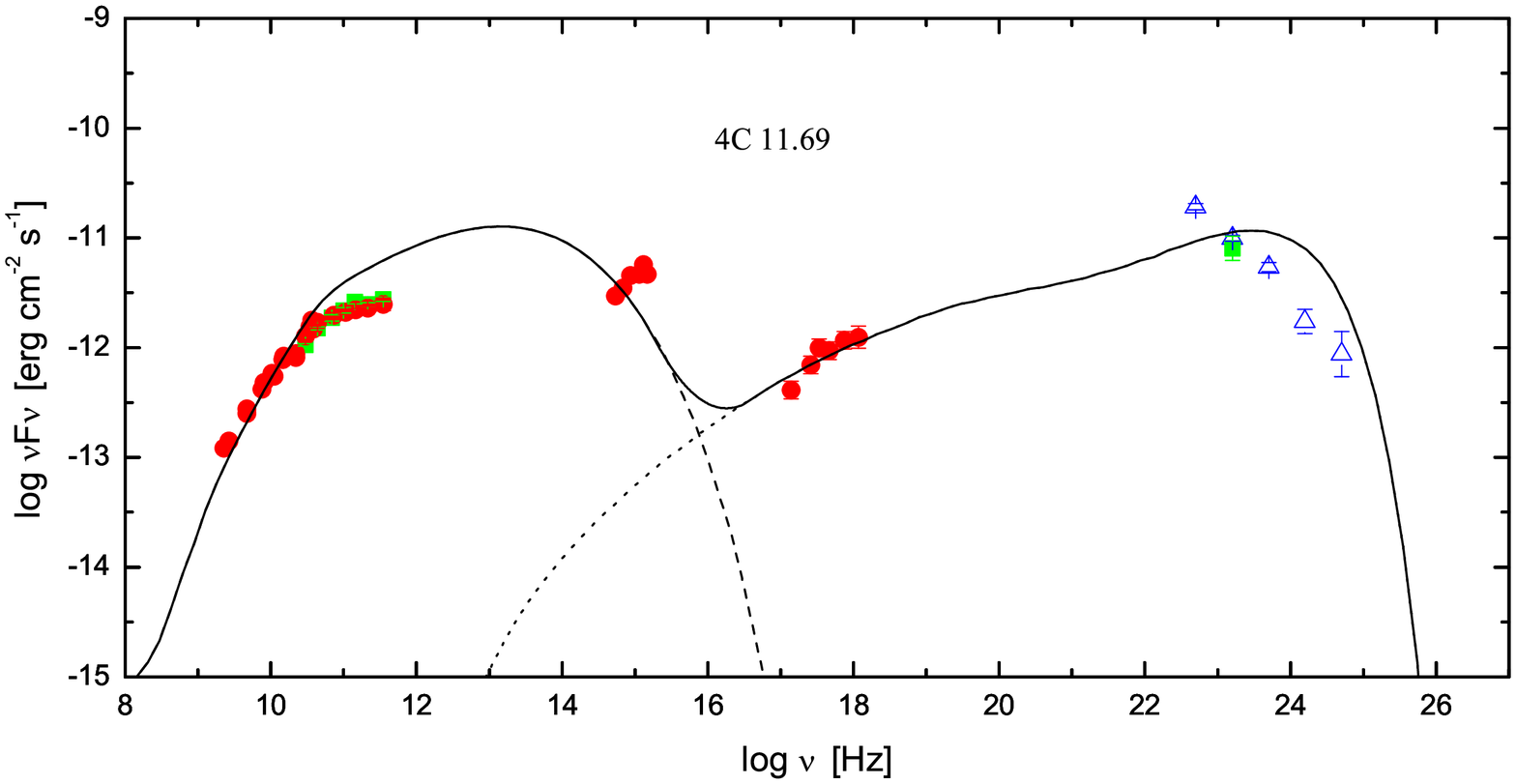}\\
        \includegraphics[scale=0.2]{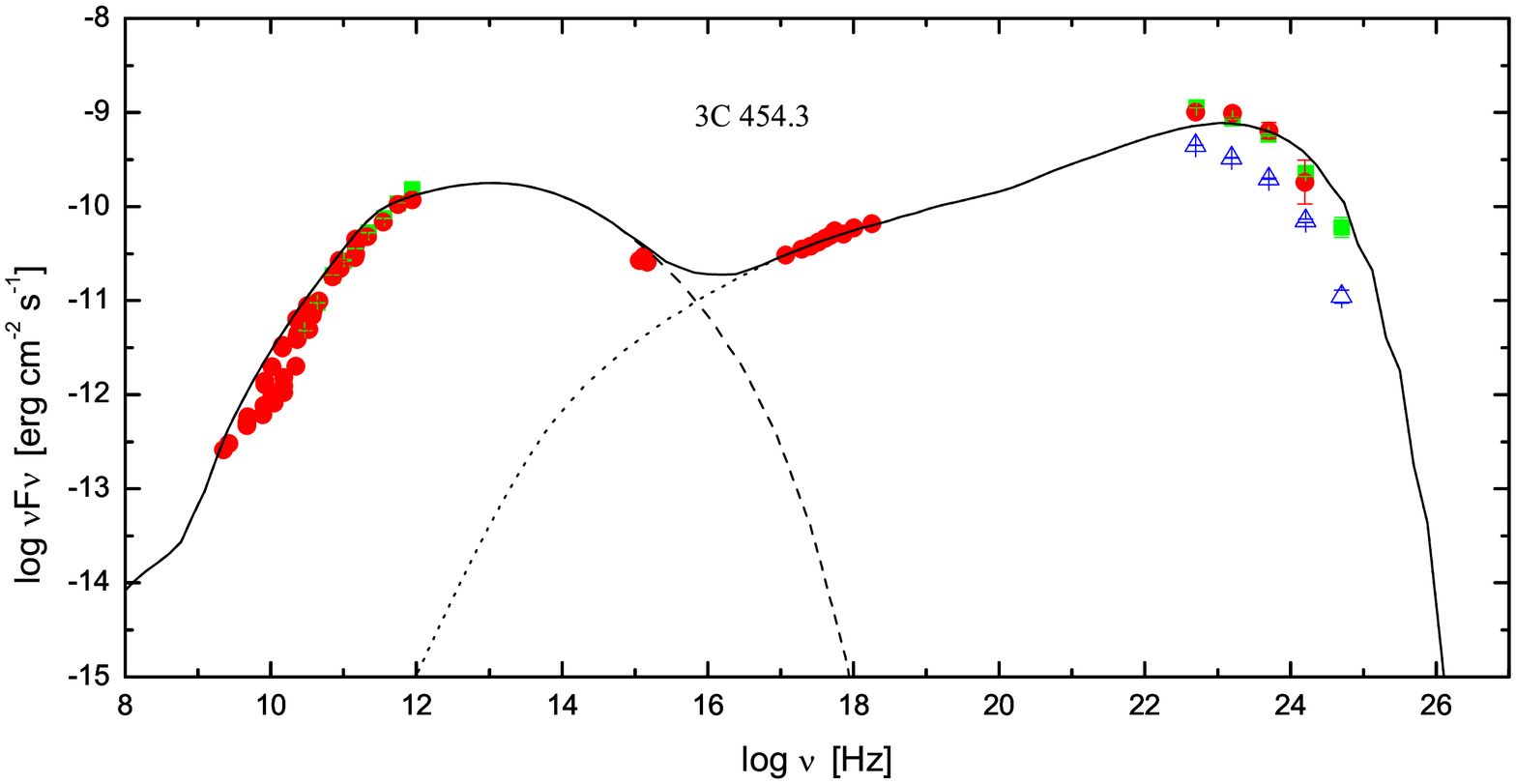}
        \includegraphics[scale=0.2]{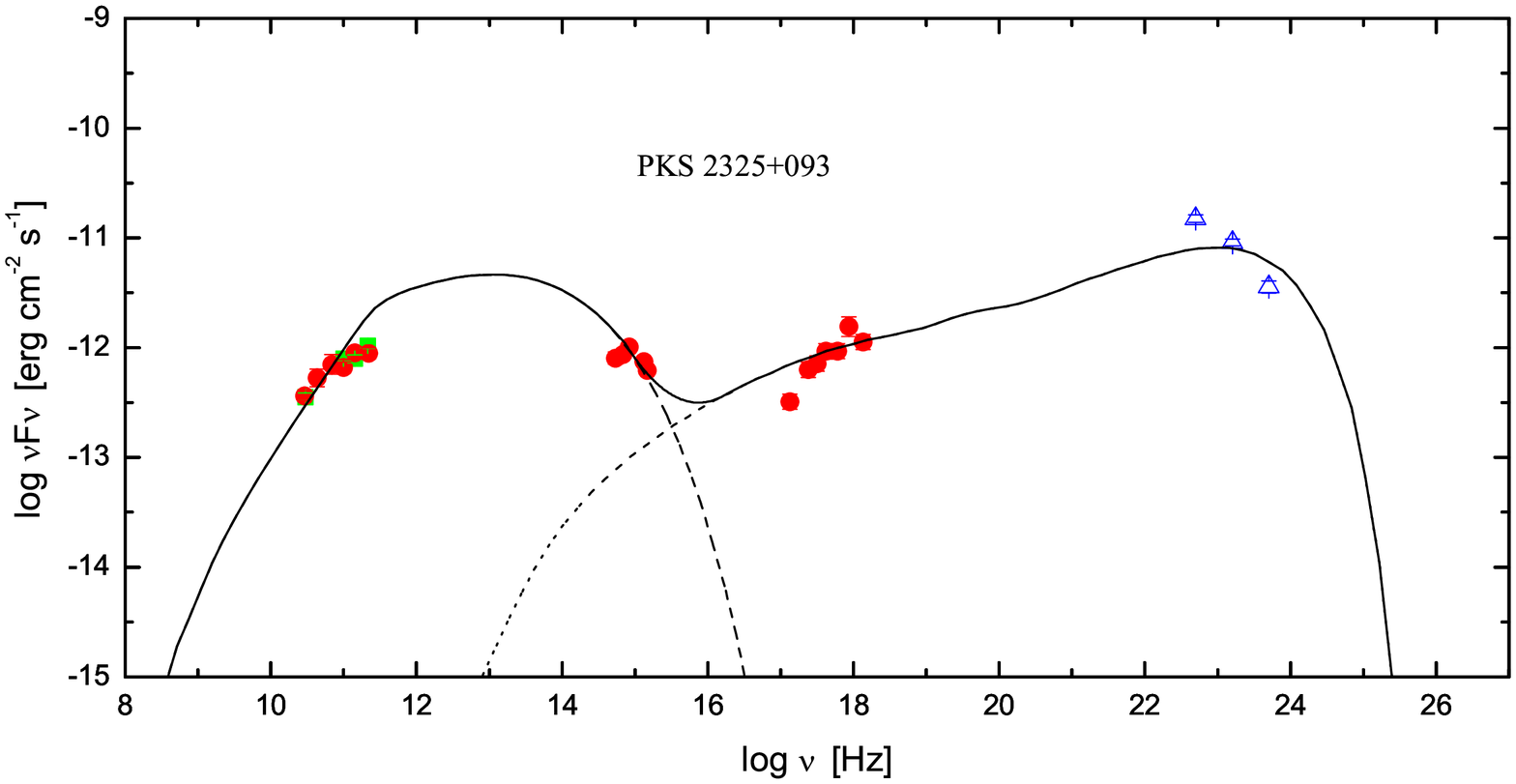}\\
    \end{tabular}
  \end{center}      		
	\caption{Comparisons of predicted multi-wavelength spectra with observed data for 4C 31.63, PKS 2204-54, PKS 2227-08, 4C 11.69, 3C 454.3, PKS 2325+093, respectively. Symbols and lines are same as in Fig. \ref{fig:3}.}
	\label{fig:8}
\end{figure*}

\subsection{Location of $\gamma$-ray emission region}
We emphasize that the characteristic distances $R_{\rm BLR}$ and $R_{\rm MT}$ could not depict the BLR and MT. In order to determine the boundaries of the BLR and MT, we take an interest in the energy density of BLR (Eq. (\ref{Eq:11})) and MT (Eq. (\ref{Eq:12})) in the jet comoving frame. It can be seen that the $u_{\rm BLR}(r)$ and $u_{\rm MT}(r)$ are decreased significantly when the distance $r$ satisfies $r>R_{\rm BLR}$, and $r>R_{\rm MT}$, respectively. The numerical results indicate that the contribution of BLR and MT could be neglected when the $u_{\rm BLR}(r)$ decreases to 10 percent of $u_{\rm BLR}$ and the $u_{\rm MT}(r)$ decreases to 1 percent of $u_{\rm MT}$, where, the $u_{\rm BLR}$ is the BLR energy density at the characteristic distance $R_{BLR}$, and the $u_{\rm MT}$ is the MT energy density at the characteristic distance $R_{MT}$, respectively. In this scenario, we could set the outer boundary of the BLR and MT at the location where the $u_{\rm BLR}(r)$ decreases to 10 percent of $u_{\rm BLR}$ and the $u_{\rm MT}(r)$ decreases to 1 percent of $u_{\rm MT}$, respectively. These results give the energy density at outer boundary $R_{\rm BLR,out}$ of the BLR $u_{\rm BLR,c}=0.1u_{\rm BLR}$ and the energy density at outer boundary $R_{\rm MT,out}$ of MT $u_{\rm MT,c}=0.01u_{\rm MT}$. Replacing these results in Eq. (\ref{Eq:11}) and Eq. (\ref{Eq:12}), we have
\begin{equation}
\frac{u_{\rm BLR,c}}{u_{\rm BLR}}=\frac{2}{1+(\frac{R_{\rm BLR,out}}{R_{\rm BLR}})^{\beta_{\rm{BLR}}}}\;,
\end{equation}
and
\begin{equation}
\frac{u_{\rm MT,c}}{u_{\rm MT}}=\frac{2}{1+(\frac{R_{\rm MT,out}}{R_{\rm MT}})^{\beta_{\rm{MT}}}}\;.
\end{equation}
Taking into account the radiation density profile $\beta_{\rm BLR}=3$ and $\beta_{\rm MT}=4$ in the model, we could deduce the outer boundary radii of the BLR as $R_{\rm BLR, out}=2.7R_{\rm BLR}$ and the MT as $R_{\rm MT, out}=4.0R_{\rm MT}$. In our model, Since we assume the external ambient fields $u_{\rm ex}(r)=u_{\rm BLR}(r)+u_{\rm MT}(r)$, we argue that the choice of the inner boundary radii of the BLR and the MT have a negligible influence.

\begin{figure}
	\centering
		\includegraphics[width=9 cm]{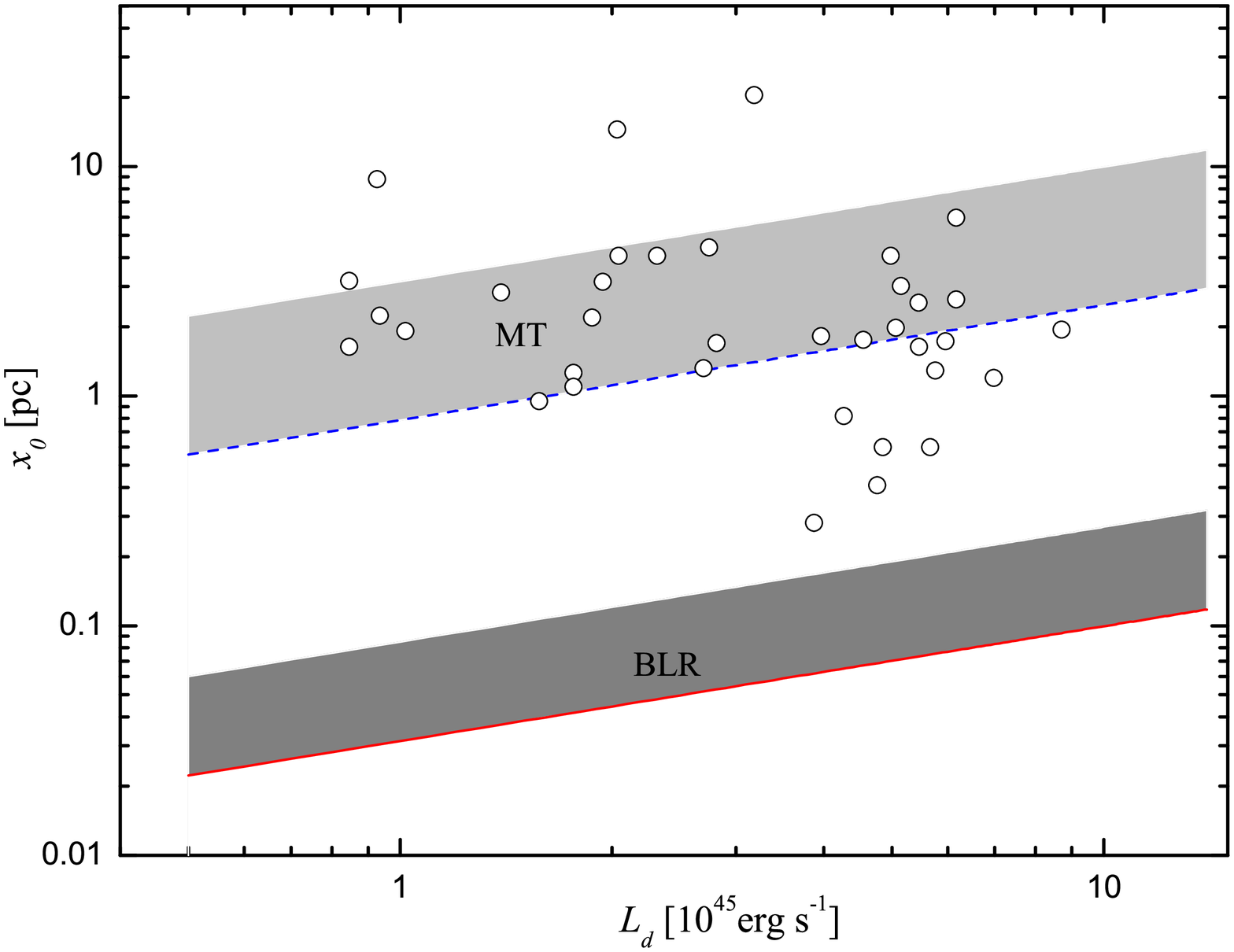}
		\caption{The location of $\gamma$-ray emission site from SMBH ($x_{0}$) as a function of the luminosity of a accretion disc ($L_{d}$). We exhibit the sample sources as open circles. The red solid and blue dashed line shows the characteristic distance of BLR and MT, respectively. The shaded regions show the extended range of BLR and MT. where are determined by $R_{BLR, out}$ and $R_{MT, out}$}
	\label{fig:9}
\end{figure}

The numerical results show that the energy dissipation region could be determined by parameter $r=x_{0}+x$ through reproducing the $\gamma$-ray spectra. It can be seen from Table {\ref{Table:1}} that modelling the SEDs of the sample give $x_{0}>$ 0.1 pc. Since the numerical results show that the MeV-GeV $\gamma$-ray emission region depends on the parameters $x_{0}$ with a constraint of $x\lesssim x_{0}$, we could use the parameter $r\sim x_{0}$ to define the radius of the $\gamma$-ray emission region, that is, the parameter $x_{0}$ could trace the location of $\gamma$-ray emission site from SMBH. Due to the characteristic distance $R_{\rm BLR}$ and $R_{\rm MT}$ being assumed to relate to the luminosity of a accretion disc $L_{d}$, we show the location of the $\gamma$-ray emission site from SMBH ($x_{0}$) as a function of the luminosity of a accretion disc ($L_{d}$) for the sample sources in Figure {\ref{fig:9}}, where the sample sources are exhibited as open circles, the red solid and blue dashed line show the characteristic distances of BLR and MT, the shaded regions show the extended ranges of BLR and MT, respectively. It can be seen that 1) the $\gamma$-ray emitting region is located at the range from 0.1 pc to 10 pc; 2) the $\gamma$-ray emitting region is located outside the BLR and within the MT; and 3) the $\gamma$-ray emitting region is located at closer to the MT range than that of the BLR.

\section{Discussion and Conclusion}
As one of the greatest challenges, the issue of the energy dissipation location for the FSRQ is essential to shed new light on the $\gamma$-ray emission mechanism (e.g. Sikora et al. 1994; Blazejowski et al. 2000; Arbeiter et al. 2002; Sokolov \& Marscher 2005; Fan et al. 2006; B$\rm \ddot{o}$ttcher 2007; Sikora et al. 2009) and/or the jet formation and collimation process (Vlahakis \& K$\rm \ddot{o}$nigl 2004; Marscher et al. 2008; Malmrose et al. 2011). In this paper, we proposed a model-dependent method to determine the location of the $\gamma$-ray emitting region. In the model, two assumptions have been made: (1) the extra-relativistic electrons are injected at the base of the jet, and then non-thermal photons are produced by both the synchrotron radiation and ICS in the energy dissipation region, where target photons dominating ICS originate from both the synchrotron photon fields and external ambient fields; and (2) the energy density of the external radiation field is a function of the distance between the position of the dissipation region and SMBH, where the energy dissipation region could be determined by the model parameter through reproducing the $\gamma$-ray spectra. We applied the model to the quasi-simultaneous multi-wavelength observed data for 36 FSRQs. Assuming a steady geometry of the jet structure and suitable physical parameters, we reproduced the multi-wavelength spectra for these 36 FSRQs, respectively.  Our results show that the $\gamma$-ray emitting regions are located at the range from 0.1 pc to 10 pc, such a range is outside the BLR and within the MT, moreover the $\gamma$-ray emitting regions are located close to the MT range than the BLR range.

The present work differs from the earlier studies on which qualitatively estimate the location of the $\gamma$-ray emitting region (e.g. Dermer et al. 2009; Ghisellini \& Tavecchio 2009; Zdziarski et al. 2012; Georganopoulos et al. 2012; Kang et al. 2015). This work focuses on modelling the distance between dissipation region and SMBH on the basis of reproducing the SEDs of a FSRQ sample. Assuming that the location of the $\gamma$-ray emitting region should be determined by the energy dissipation or particle acceleration processes within the relativistic jet, we could quantify the location of the $\gamma$-ray emitting region. Despite a large number of chanciness in our results, we argue that the model presents a diagnostic on the location of the $\gamma$-ray emission region.

A potential drawback of the model-dependent diagnostic is that our results significantly depend on the boundaries of the BLR and MT. Though we could estimate the characteristic distance of the BLR and MT (Bentz et al. 2006; Kaspi et al. 2007; Ghisellini \& Tavecchio 2008; Bentz et al. 2009), the widths of the BLR shell and MT shell are poorly known. In order to estimate the $\gamma\gamma$ absorption in the BLR, B$\rm \ddot{o}$ttcher \& Paul (2016) adopt a thin BLR shell with $R_{\rm BLR,~in}=0.9R_{\rm BLR}$ and $R_{\rm BLR,~out}=1.1R_{\rm BLR}$. It is likely that the MT is a clumpy structure with a range of radii extending close to the SMBH, where the temperature is just blow dust sublimation (Nenkova et al. 2008a; 2008b). Following this scenario, Malmrose et al. (2011) argued that the inner radius of the MT is located in 1-2 pc. In the present work, we used a simple numerical constraint with the $u_{\rm BLR}(r)$ decreasing to 10 percent of $u_{\rm BLR}$ and the $u_{\rm MT}(r)$ decreasing to 1 percent of $u_{\rm MT}$ to determine the outer boundary of the BLR and MT, respectively. On the basis of the simple numerical constraint, a direct evidence for the \emph{far site} scenario could be obtained.

It is believed that in the \emph{far site} scenario the combined effects of the decrease of the scattering cross section (Tavecchio \& Ghisellini 2008)and of the possible $\gamma\gamma$ absorption (e.g. Donea \& Protheroe 2003; Liu et al. 2008b) would result in a spectra cut-off at large photon energy  around 1 TeV. However, spectral breaks at a few GeV have been found in several FSRQs (e.g. Abdo et al. 2009b; Abdo et al. 2010; Ackermann et al. 2010; Poutanen \& Stern 2010; Abdo et al. 2011; Stern \& Poutanen 2011). We note that if we adopt a relaxed constraint with the $u_{\rm BLR}(r)$ decreasing to 1 percent $u_{\rm BLR}$ and the $u_{\rm MT}(r)$ decreasing to 0.1 percent of $u_{\rm MT}$, the $\gamma$-ray emitting regions of the sources PKS 0454-234, PKS 1124-186, 3C 279, and 3C 454.3 could be embedded into the BLR, and the sources \uppercase\expandafter{\romannumeral3} ZW 2, PKS 0420-01, and 4C 06.69 could be embedded into the MT.

We note that for the \emph{far site} scenario the $\gamma$-ray emitting region is possible located at outside the BLR and the external ambient fields for ICS should be dominated by the MT. This result is consistent with the earlier issue that is deduced from the seed factor (SF) (Georganopoulos et al. 2012). On the other hand, there are several observational evidences to sustain our conclusion, such as exploring the velocity field distributions of the jet in M87. Asada et al. (2014) suggested that most of the jet energy is dissipated at distances of tens of pc from the central black hole; By a probability, the FSRQs 3C 279 (Aleksic et al. 2011a), 4C 21.35 (Aleksic et al. 2011b), PKS 1510-089 (Hauser et al. 2011) have been detected by Cherenkov telescopes in the sub-TeV energy range; The correlation between the millimeter variability and the $\gamma$-ray light curves indicates that the $\gamma$-ray emitting region should be located at 14 pc (Agudo et al. 2011a) and $\sim12$ pc (Agudo et al. 2011b) from SMBH in the jet of OJ 287.

The key idea of the method is to evolve the energy dissipation region from the base to the end of the jet. Since more than one component, located at different regions from the SMBH, are simultaneously active, we could expect a energy dependent $\gamma$-ray emitting region. It is considered that the inner regions, emitting within the BLR, could contribute to the $\gamma$-rays component below 10 GeV, while the outer regions, beyond the BLR,  could contribute to the $\gamma$-rays component above 10 GeV (e.g. Tavecchio et al. 2010; Liu et al. 2011a). However, in the context model frame, we found that the higher energy photons are produced in the inner dissipation region, and the lower energy photons are produced in the external dissipation region. Therefore we argue that a special accelerated mechanism (Guo et al. 2015; 2016; Levinson \& Globus 2016) should be taken into account for the verification these issues. We leave these possibility to our future work.


\acknowledgments
\section*{acknowledgments}
We thank the anonymous referee for valuable comments and suggestions. This work is partially supported by the National Natural Science Foundation of China under grants 11463007, 11573060, 11673060, the Strategic Priority Research Program ``the Emergence of Cosmological Structures" of the Chinese Academy of Sciences (Grant XDB09000000), Science and Technology in support of Yunnan Province Talent under grants 2012HB014, and the Natural Science Foundation of Yunnan Province under grant 2013FD014, 2016FB003. This work is also supported by the Key Laboratory of Particle Astrophysics of Yunnan Province (grant 2015DG035).

\clearpage

\end{document}